\documentclass[%
 reprint,
 superscriptaddress,
 preprintnumbers,
 amsmath,amssymb,
 aps,
 superscriptaddress,
 prb,
 floatfix,
 twocolumn,
 longbibliography,
 showpacs
]{revtex4-2}

\usepackage[version=3]{mhchem} 
\usepackage{lmodern}
\usepackage{color}
\usepackage[dvipsnames,svgnames,table]{xcolor}
\usepackage[colorlinks=true,linkcolor=blue,urlcolor=blue,citecolor=blue]{hyperref}
\usepackage{graphicx}
\usepackage{dcolumn}
\usepackage{array}
\usepackage{bm}
\usepackage{subfigure}
\usepackage{amssymb}
\usepackage{multirow}
\usepackage{tabularx}
\usepackage{amsmath}
\usepackage{bbm}
\usepackage{makecell}

\renewcommand{\vec}[1]{\boldsymbol{#1}}
\newcommand{\mat}[1]{\mathbf{#1}}
\DeclareMathSymbol{\shortminus}{\mathbin}{AMSa}{"39}

\newcommand{\order}[1]{\mathcal{O}\left(#1\right)}

\newcommand{\etal}{\textit{et~al.}}

\DeclareMathOperator*{\sumint}{%
\mathchoice%
  {\ooalign{$\displaystyle\sum$\cr\hidewidth$\displaystyle\int$\hidewidth\cr}}
  {\ooalign{\raisebox{.14\height}{\scalebox{.7}{$\textstyle\sum$}}\cr\hidewidth$\textstyle\int$\hidewidth\cr}}
  {\ooalign{\raisebox{.2\height}{\scalebox{.6}{$\scriptstyle\sum$}}\cr$\scriptstyle\int$\cr}}
  {\ooalign{\raisebox{.2\height}{\scalebox{.6}{$\scriptstyle\sum$}}\cr$\scriptstyle\int$\cr}}
}

\iffalse
    \newcommand{\todo}[1]{{\color{red}[#1]}} 
    \newcommand{\question}[1]{{\color{blue}[#1]}}
    \newcommand{\reply}[1]{{\color{magenta}[#1]}}
\else
    \newcommand{\todo}[1]{}               
    \newcommand{\question}[1]{}
    \newcommand{\reply}[1]{}
\fi

\begin{document}


\title{Reweighting Estimators for Density Response in Path Integral Monte Carlo: Applications to linear, nonlinear and cross-species density response}

\author{Pontus Svensson}
\email{p.svensson@hzdr.de}
\affiliation{Institute of Radiation Physics, Helmholtz-Zentrum Dresden-Rossendorf (HZDR), D-01328 Dresden, Germany}

\author{Thomas~Chuna}
\affiliation{Center for Advanced Systems Understanding (CASUS), Helmholtz-Zentrum Dresden-Rossendorf (HZDR), D-02826 G\"orlitz, Germany}

\author{Jan Vorberger}
\affiliation{Institute of Radiation Physics, Helmholtz-Zentrum Dresden-Rossendorf (HZDR), D-01328 Dresden, Germany}

\author{Zhandos~A.~Moldabekov}
\affiliation{Institute of Radiation Physics, Helmholtz-Zentrum Dresden-Rossendorf (HZDR), D-01328 Dresden, Germany}

\author{Paul~Hamann}
\affiliation{Institute of Radiation Physics, Helmholtz-Zentrum Dresden-Rossendorf (HZDR), D-01328 Dresden, Germany}
\affiliation{Institut für Physik, Universität Rostock, D-18057 Rostock, Germany}

\author{Sebastian~Schwalbe}
\affiliation{Institute of Radiation Physics, Helmholtz-Zentrum Dresden-Rossendorf (HZDR), D-01328 Dresden, Germany}

\author{Panagiotis~Tolias}
\affiliation{Electromagnetics and Plasma Physics, Royal Institute of Technology (KTH), Stockholm, SE-100 44, Sweden}

\author{Tobias~Dornheim}
\affiliation{Institute of Radiation Physics, Helmholtz-Zentrum Dresden-Rossendorf (HZDR), D-01328 Dresden, Germany}
\affiliation{Center for Advanced Systems Understanding (CASUS), Helmholtz-Zentrum Dresden-Rossendorf (HZDR), D-02826 G\"orlitz, Germany}

\date{\today}

\begin{abstract}
    We present density response estimators for Monte Carlo simulations that are based on a reweighting procedure, where the samples of an unperturbed system are used to estimate the properties of a system perturbed by an external harmonic potential. This allows the linear and nonlinear static density response to be estimated purely from simulations of the unperturbed system. The method is demonstrated for the uniform electron gas under warm dense matter and strongly coupled conditions using \textit{ab initio} path integral Monte Carlo simulations. The performance of the method with respect to the number of particles and the number of imaginary time slices is investigated. The scheme is generalised to consider multiple external perturbations, acting on different species and with different wavenumbers, giving one access to additional cross-species density response functions and the complete quadratic response function resolved for both wave number arguments through mode coupling. The flexibility of the methodology opens the possibility to investigate numerous new density response properties to further advance our understanding of interacting quantum many-body systems across a broad range of applications.
\end{abstract}
\maketitle

\section{Introduction}\label{sec:intro}

Exact analytical solutions for interacting quantum many-body problems are exceedingly rare, with only a few known examples. Therefore, our understanding of most systems has historically been largely based on perturbative treatments, yielding quantitative descriptions of weakly interacting systems and to some extent simple metals~\cite{brovman1974phonons,louis1998extending}. Nowadays, numerical computations go beyond perturbative treatments, but the behaviour of perturbations around such numerical approximations is still key to understanding various system properties, covering a broad range of conditions from ultracold atoms~\cite{filinov2010berezinskii,filinov2012collective,dornheim2022path} to exotic quark-gluon plasmas~\cite{filinov2013color}. A particular example is the density response of a system due to an external scalar potential, which can be directly related to scattering~\cite{sturm1993dynamic,glenzer2009x}, effective particle interactions~\cite{nagao2003enhanced}, stopping power~\cite{ritchie1959interaction}, etc.

A particularly interesting physical regime is given by so-called \textit{warm dense matter} (WDM)~\cite{vorberger2025roadmapwarmdensematter,graziani2014frontiers}, which is a complex state found in astrophysical objects such as large planets~\cite{guillot1999interiors,fortney2010interior,helled2020understanding,french2012ab,preising2023material}, stars~\cite{fortov2009extreme,kippenhahn2012stellar,chabrier2000cooling}, and the atmospheres of neutron stars~\cite{gudmundsson1983structure,haensel2007neutron}. States of WDM are also created in human-made applications such as inertial confinement fusion (ICF) experiments~\cite{nuckolls1972laser,betti2016inertial,hurricane2023physics} and in the synthesis of novel materials~\cite{miao2020chemistry}. In WDM, electrons are characterised by $r_s$ -- the ratio between the Wigner-Seitz radius and the Bohr radius -- and $\Theta$ -- the ratio of the thermal excitation energy and the electronic Fermi energy -- both being of order unity, such that interactions, quantum degeneracy, and thermal excitations are simultaneously substantial~\cite{bonitz2020ab,bonitz2024principles,vorberger2025roadmapwarmdensematter,ott2018recent}. The lack of a small expansion parameter in WDM systems requires the use of nonperturbative methods, but perturbative treatments are still central in constraining aspects of computational models, e.g., the exchange-correlation functional in density functional theory (DFT)~\cite{perdew2008restoring,tao2008nonempirical,sun2015strongly,pribram2016thermal} or the kinetic energy functionals in orbital-free DFT~\cite{wang1992kinetic,xu2019nonlocal,moldabekov2023imposing}. 

Recently, a large number of computational studies have been carried out for the density response and directly related properties under WDM conditions, using semiclassical molecular dynamics (MD)~\cite{larder2019fast,kahlert2026structural}, wave packet MD~\cite{davis2020ion,svensson2024modeling,svensson2025modeling}, DFT~\cite{moldabekov2022density}, DFT-MD~\cite{moldabekov2022density,schorner2022extending}, time-dependent DFT~\cite{baczewski2016x,moldabekov2023linear,hentschel2025statistical,moldabekov2025applying,moldabekov2025enhancing} and path integral Monte Carlo (PIMC)~\cite{dornheim2018ab,dornheim2020nonlinear,bohme2022static,dornheim2024ab}. The outcome of these investigations has directly led to new descriptions of dynamic structure factors~\cite{dornheim2020effective,schorner2023x}, refinements of density estimations from x-ray Thomson scattering measurements~\cite{dornheim2025unraveling,schwalbe2025static,poole2024multimessenger} and stringent tests of DFT functionals~\cite{moldabekov2021relevance,moldabekov2022benchmarking,moldabekov2023assessing,moldabekov2025density}.

Computational studies of static density response functions typically either evaluate the complete response of the system to an external potential energy perturbation or evaluate the multi-particle density correlation functions of the unperturbed system. The direct perturbation approach computes the density response~\cite{sakko2010time,dalton2013linear,dornheim2020nonlinear} or related properties~\cite{moroni1992static,sugiyama1992static,bowen1994static,moroni1995static}, being straightforwardly based on the functional derivative definition of density response functions~\cite{hu1988z,bergara1999quadratic,dornheim2021density,vorberger2025green}. The method requires multiple simulations at various perturbation strengths to extract the arbitrary-order response coefficients, and multiple simulations at many perturbation wavelengths to acquire the full spectrum of wavenumber-resolved static responses. The correlation function approach, on the other hand, allows for the simultaneous evaluation of multiple response coefficients, but it requires response theory to connect the arbitrary-order density correlation functions at thermodynamic equilibrium to the arbitrary-order density response functions~\cite{kubo1966fluctuation,hansen1993theory,pines2018theory,sugiyama1992static,groth2019ab}. This connection becomes increasingly challenging both to establish and to evaluate when considering higher order responses and multi-component systems. A particularly elusive example concerns the cubic response at the first harmonic~\cite{vorberger2025green}, which so far has only been evaluated using the direct perturbation approach and formal functional derivatives of the free energy functional~\cite{moldabekov2025generalized}. Furthermore, in methods such as restricted path integral Monte Carlo (RPIMC), the correlation function approach based on the imaginary time correlation function (ITCF)~\cite{dornheim2021nonlinear} is not available as the nodal restriction breaks imaginary-time translation invariance~\cite{ceperley1991fermion}. This makes it highly desirable to find an alternative approach without the requirement of performing a large number of perturbed simulations.

In this paper, we introduce a simulation method that combines the advantages of the direct perturbation approach (immediate access to linear and nonlinear density response) and the correlation function approach (a single simulation suffices for all the full wavenumber spectrum). The method is based on the same underlying theory as the direct perturbation approach. However, instead of introducing a perturbation to the system, a reweighting methodology is employed that allows the measurement of the properties of one system based on the reweighted samples from another similar system. In this manner, the reweighting estimator approach allows for the acquisition of the complete density response from a single simulation within the perturbation formulation. The method is used to derive linear and nonlinear density response function estimators, both within the same and across species. The presented method stays close to the definition of the density response, but overcomes the computational bottleneck of performing separate simulations for individual wavenumbers and perturbation amplitudes. The effortless interpretation of the method will allow us to extend the formulation to extract previously inaccessible response functions. In particular, this is possible by considering multiple fictitious perturbations simultaneously. First, we show how to access all species combinations for the higher order response functions, and second, the complete spectrum of quadratic response functions resolved over two wave vector arguments. As a practical application, we simulate the finite temperature uniform electron gas (UEG)~\cite{giuliani2008quantum,dornheim2018uniform}, which is often considered the canonical system of interacting electrons.

The remainder of this manuscript is organised as follows. Section \ref{sec:diag_response_estimators} presents the general idea of reweighting and the theoretical background required to evaluate the density response using this formulation. This new estimator is used in Section \ref{sec:single_pert} for the uniform electron gas (UEG). The discretisation error is thoroughly investigated, and a nonlinear extension of the density stiffness theorem is tested against simulation results. In Section \ref{sec:mult_pert_species}, an estimator that fictitiously perturbs two species simultaneously is introduced and the  extraction of additional cross-species density response coefficients is demonstrated. Section \ref{sec:mult_pert_q} presents and demonstrates a scheme based on two fictitious perturbations to explore the complete spectrum of quadratic perturbations. The conclusions of the manuscript are summarised in Section \ref{sec:conclusions}.

\section{Reweighting estimators for density response}\label{sec:diag_response_estimators}

In quantum Monte Carlo simulations, importance sampling is used to efficiently evaluate high dimensional integrals~\cite{landau2021guide}. Reweighting is a method to extract additional information from existing Monte Carlo samples based on the idea that if two systems are sufficiently similar, then a good set of samples for one system can be reweighted to sample the other system. McDonald \& Singe used similar techniques already in 1967 to investigate the thermodynamic properties of liquid Argon~\cite{mcdonald1967calculation}, and more recently related ideas have been employed to study the system behaviour close to phase transitions~\cite{ferrenberg1988new}, to estimate the energy of similar wave functions in variational Monte Carlo~\cite{ceperley1979quantum}, to estimate differences between closely related systems in diffusion Monte Carlo~\cite{boronat1999quantum}, to consider multiple particle statistics simultaneously in PIMC~\cite{dornheim2025reweighting}, and to improve the efficiency of grand canonical PIMC simulations~\cite{hamann2026reweighting}. We will extend this method for the evaluation of density response functions in PIMC.

\subsection{Reweighting estimators}\label{sec:diag_response_estimators_estimator}

The statistical average of an observable $\Hat{O}$ in a system $a$ is
\begin{equation}
    \langle \Hat{O} \rangle_a = \frac{1}{Z_a} \sumint d\vec{X}\; W_a(\vec{X}) O(\vec{X}),
    \label{eq:def_average}
\end{equation}
where the integration is performed over the space of many-body configurations $\vec{X}$, $W_a(\vec{X})$ is the weight of configuration $\vec{X}$ and 
\begin{equation}
    Z_a = \sumint d\vec{X}\; W_a(\vec{X})
\end{equation}
is the partition function that acts as the normalisation. 

Consider the average of an observable $\Hat{O}$ in two systems $a$ and $b$, respectively. By introducing the rescaled observable $O_{ab}(\vec{X}) = W_b(\vec{X}) O(\vec{X}) / W_{a}(\vec{X})$, the average in system $b$ can be reformulated as a combination of averages in system $a$,
\begin{equation}
    \begin{aligned}
        \langle \Hat{O} \rangle_b &= \frac{Z_a}{Z_b} \frac{1}{Z_a} \sumint d\vec{X}\; W_a(\vec{X}) \underbrace{\frac{W_b(\vec{X})}{W_a(\vec{X})} O(\vec{X})}_{O_{ab}(\mathbf{X})}\\
        &= \frac{\langle \Hat{O}_{ab} \rangle_a}{\langle \Hat{1}_{ab} \rangle_a},
    \end{aligned}
    \label{eq:generla_reweighting}
\end{equation}
where $\langle \Hat{1}_{ab} \rangle_a = Z_b / Z_a$ which can be computed as the average of the reweighted identity operator. In the second equality, it has been assumed that the phase space on which $W_b(\vec{X})$ is non-zero is a subset of or equal to the phase space on which $W_a(\vec{X})$ is non-zero. The possibility of evaluating an average in system $b$ purely from samples of system $a$ will allow us to evaluate observables in a large number of systems $b$ similar to $a$ in parallel.

Although reweighting saves on computational work by reusing existing samples, some of the limitations of the methods can be understood in terms of Eq.~\eqref{eq:generla_reweighting}. The denominator is a ratio of partition functions and can be formulated in terms of a free energy difference
\begin{equation}
    \langle \Hat{1}_{ab} \rangle_a = e^{-\beta (F_b - F_a)},
    \label{eq:denomenator_free_energy}
\end{equation}
where $F_a$ and $F_b$ are the free energy of system $a$ and $b$, respectively. A common inverse temperature $\beta=(k_\text{B}T)^{-1}$ has been used for the two systems, but it is not required. When the difference between the two systems is described by a localised change, e.g., a localised external perturbation or the property of a single particle is changed, the free energy difference is expected to be close to a constant with respect to system size. On the other hand, when the change is global in nature, e.g., the interaction between particles is altered or an extensive external perturbation is applied, the leading order change in free energy is expected to be proportional to the number of particles $N$. The result is an exponentially decaying or increasing denominator with particle number, and the numerator will have a similar trend. Conceptually, a denominator far from unity means that the two systems are substantially different for the purpose of sampling. Therefore, the resulting structure of the reweighting scheme shares some of the challenging features of the fermion sign problem in PIMC~\cite{hatano1994data}, as a ratio of averages of two highly correlated properties must be resolved, something which will become increasingly challenging for larger system sizes when the difference between the systems is global.

\subsection{Reweighting in fermionic PIMC}

In PIMC with Fermi statistics, the weight function $W_a(\vec{X})$ can be negative and cannot be considered as a probability distribution. Instead, a corresponding system with Bose statistics is sampled, resulting in the fermion sign problem~\cite{loh1990sign,troyer2005computational,dornheim2018uniform,dornheim2019fermion}. Concretely, estimators take the form
\begin{equation}
    \langle \Hat{O} \rangle_a = \frac{ \langle \Hat{S} \Hat{O} \rangle_a' }{ \langle \Hat{S} \rangle_a'}, 
    \label{eq:fermionic_estimator}
\end{equation}
with $\langle \cdot \rangle_a'$ the average in system $a$ with Bose statistics and $\Hat{S}$ the sign operator. This approach is a reweighting from a system with Bose statistics to one with Fermi statistics, as $\Hat{S}$ is the ratio of fermionic and bosonic weights.

When performing the reweighting in fermionic PIMC, two types of reweighting are performed together. Combining Eqs.~\eqref{eq:generla_reweighting} and \eqref{eq:fermionic_estimator} results in the final estimator
\begin{equation}
    \langle \Hat{O} \rangle_b = \frac{\langle \Hat{S} \Hat{O}_{ab} \rangle_a'}{\langle \Hat{S} \Hat{1}_{ab} \rangle_a'},
\end{equation}
where the average sign $\langle \Hat{S} \rangle_a'$ cancels. The reweighting factor $\Hat{S} \Hat{1}_{ab}$ reweights from system $a$ with Bose statistics to system $b$ with Fermi statistics. Therefore, the denominator still has properties similar to those of the average sign, where
\begin{equation}
    \langle \Hat{S} \Hat{1}_{ab} \rangle_a' = e^{-\beta (F_b - F_a')}
\end{equation}
and $F_a'$ is the free energy of system $a$ with Bose statistics. The sign problem for the PIMC reweighting estimators takes this alternative form, but overall it shares similar characteristics.

\subsection{Density response}

The induced density perturbation $\langle \Hat{n}_{\text{ind}}^{\Bar{c}}(\Bar{\vec{r}}) \rangle$ in species $\Bar{c}$ due to external potentials $V_{\text{ext}}^{c_i}(\vec{r}_i)$ acting on species $c_i$ is~\footnote{In Ref.~\cite{vorberger2025green}, the irst four response functions are denoted as $L(\vec{k}_1)$, $\frac{1}{2}Y(\vec{k}_1, \vec{k}_2)$, $\frac{1}{6}Z(\vec{k}_1, \vec{k}_2, \vec{k}_3)$ and $\frac{1}{4!}W^{(4)}(\vec{k}_1, \vec{k}_2, \vec{k}_3, \vec{k}_4)$, respectively. The arbitrary order response functions where also defined analogously in Ref.~\cite{tolias2023unravelling}. This is the generalisation of the formulation in Ref.~\cite{dornheim2021nonlinear} where the external potential acts has a dependence on the species.}
\begin{widetext}
\begin{equation}
    \langle \Hat{n}_{\text{ind}}^{\Bar{c}}(\Bar{\vec{r}}) \rangle = \sum_{n = 1}^{\infty} \sum_{c_1} \cdots \sum_{c_n} \int d\vec{r}_1 \cdots d\vec{r}_n\;  \chi^{(n)}_{\Bar{c}c_1\cdots c_n}(\Bar{\vec{r}}, \vec{r}_1, \dots, \vec{r}_n) \;\; V_{\text{ext}}^{c_1}(\vec{r}_1) \cdots V_{\text{ext}}^{c_n}(\vec{r}_n),
\end{equation}
\end{widetext}
where the sums over $c_i$ are taken over all species in the system. Thus, the species resolved $n$th order density response function is formally defined in terms of functional derivatives of the induced density perterbation: 
\begin{equation}
    \chi^{(n)}_{\Bar{c}c_1\cdots c_n}(\Bar{\vec{r}}, \vec{r}_1, \dots, \vec{r}_n) = \frac{1}{n!} \frac{ \delta^{n} \langle \Hat{n}_{\text{ind}}^{\Bar{c}}(\Bar{\vec{r}}) \rangle }{\delta V_{\text{ext}}^{c_1}(\vec{r}_1) \cdots \delta V_{\text{ext}}^{c_n}(\vec{r}_n) }.
\end{equation}
For uniform systems with translation invariance, the response function can only depend on the separations, and the number of position arguments can be reduced by $\chi^{(n)}_{\Bar{c}c_1\cdots c_n}(\Bar{\vec{r}}, \vec{r}_1, \dots, \vec{r}_n) = \chi^{(n)}_{\Bar{c}c_1\cdots c_n}(\vec{r}_1 - \Bar{\vec{r}}, \dots, \vec{r}_n - \Bar{\vec{r}})$. 

When perturbing a system with an external harmonic potential, density response properties can be investigated. Let system $a\equiv0$ be the unperturbed system, and $b\equiv(\vec{q}, A, c)$ be the same system where the species $c$ is harmonically perturbed. The external potential is
\begin{equation}
    \Hat{V}_{\text{ext}} = 2A \int\! d\vec{r}\; \cos(\vec{q} \cdot \vec{r}) \Hat{n}^{c}(\vec{r}),
    \label{eq:perterbation}
\end{equation}
with $\Hat{n}^{c}(\vec{r})$ the microscopic density operator for species $c$ and with $A$ the perturbation amplitude. The density response of the species $\Bar{c}$ is measured through the observable $\Hat{O} = \Hat{n}_{\vec{k}}^{\Bar{c}} / \Omega = \Hat{\rho}^{\Bar{c}}_{\vec{k}}$, with $\Hat{n}_{\vec{k}}^{\Bar{c}}$ the Fourier transformed microscopic density operator and with $\Omega$ the volume of the simulation cell. This response will be measured with the reweighting estimator approach in a PIMC simulation within the primitive factorisation with $P$ imaginary time slices and $N_{c}$ particles of species $c$. In this case, the reweighting factor is
\begin{equation}
    \frac{W_{\vec{q}, A, c}(\vec{X})}{W_0(\vec{X})} = \exp\left( -2A\epsilon \sum_{\alpha = 0}^{P - 1} \sum_{j = 1}^{N_c} \cos(\vec{q} \cdot \vec{r}_{j,\alpha}) \right),
    \label{eq:reweighting_factor}
\end{equation}
where $\vec{r}_{j,\alpha}$ is the position of particle $j$ on imaginary time slice $\alpha$ and with $\epsilon = \beta/P$. Since bosonic weights are always non-zero in PIMC, the integration space in Eq.~\eqref{eq:def_average} is the same for the unperturbed and perturbed systems.

For originally uniform systems and with the external harmonic perturbation described by Eq.~\eqref{eq:perterbation}, the system exclusively responds at $\vec{k}$-vectors that are integer multiples of the perturbation wavevector $\vec{q}$. Specifically:~\cite{tolias2023unravelling}
\begin{align}
    \langle \Hat{\rho}^{\Bar{c}}_{m\vec{q}} \rangle_{\vec{q}, A, c} &= \sum_{l = 0}^{\infty} \chi_{\Bar{c}c}^{(m,m+2l)}(\vec{q}) A^{m + 2l}.
    \label{eq:response_singel_pert}
\end{align}
In the above, $\chi_{\Bar{c}c}^{(m, n)}$ represents the $n$th order response of species $\Bar{c}$ at the wave vector $\vec{k} = m \vec{q}$ ($m$th harmonic). These specific structures can be understood due to mode coupling, where lower order perturbations act as sources for higher order ones~\cite{moldabekov2025generalized}.

The coefficients $\chi_{\Bar{c}c}^{(m, n)}(\vec{q})$ are directly related to the response functions. The leading order (or diagonal $n=m$) coefficients for the first three harmonics are~\cite{dornheim2021density,tolias2023unravelling}
\begin{subequations}
\begin{align}
        \chi^{(1,1)}_{\Bar{c}c}(\vec{q}) &= \chi^{(1)}_{\Bar{c}c}(\vec{q}),\\
        \chi^{(2,2)}_{\Bar{c}c}(\vec{q}) &= \chi^{(2)}_{\Bar{c}cc}(\vec{q}, \vec{q}),\\
        \chi^{(3,3)}_{\Bar{c}c}(\vec{q}) &= \chi^{(3)}_{\Bar{c}ccc}(\vec{q}, \vec{q}, \vec{q}),
\end{align}
where $\chi^{(n)}_{\Bar{c}c_1\cdots c_n}(\vec{k}_1, \dots, \vec{k}_n)$ is the Fourier transform of $\chi^{(n)}_{\Bar{c}c_1\cdots c_n}(\vec{r}_1, \dots, \vec{r}_n)$. The higher order (or off-diagonal $n>m$) coefficients are related to static density response functions that feature other $\vec{k}$-vector combinations as arguments, e.g., for the cubic response at the first harmonic~\cite{dornheim2021density,tolias2023unravelling}
\begin{equation}
    \chi^{(1, 3)}_{\Bar{c}c}(\vec{q}) = 3\Bar{\chi}^{(3)}_{\Bar{c}ccc}(-\vec{q}, \vec{q}, \vec{q}),
    \label{eq:third_first_harmonic}
\end{equation}
or for the quartic response at the second harmonic~\cite{tolias2023unravelling}
\begin{equation}
    \begin{aligned}
        \chi^{(2, 4)}_{\Bar{c}c}(\vec{q}) = 4\Bar{\chi}^{(4)}_{\Bar{c}cccc}(-\vec{q}, \vec{q}, \vec{q}, \vec{q}).
    \end{aligned}
    \label{eq:fourth_second_harmonic}
\end{equation}%
\label{eq:coefficent_definition_single_perterbation}%
\end{subequations}
In the above, we have included an additional ``bar'' over the response function to highlight that we consider a symmetrised description of the response functions in terms of their wave number arguments~\cite{sitenko1982fluctuations}. In general, there is some freedom in the definition of the nonlinear response functions, but the observable quantity is the average over wavevector permutations, and there is no limitation in considering a symmetrised description. Therefore, a single harmonic perturbation gives access to the full linear response function and higher order response functions along certain cuts in $\vec{k}$-space and for specific species combinations. In Sections \ref{sec:species_estimator} and  \ref{sec:complete_response_estimators}, we will address how to access all species combinations and the complete $\vec{k}$-vector space for the quadratic response, respectively. However, the reweighting procedure is straightforward to generalise to such settings. 

The density perturbation is computed for various perturbation amplitudes $A$ using the reweighting. To extract the response coefficients from the reweighting estimates of $\langle \Hat{\rho}_{\vec{k}}^{\Bar{c}} \rangle_{\vec{q},A,c}$, a polynomial fitting is carried out according to Eq.~\eqref{eq:response_singel_pert}. The amplitude strength where the reweighting scheme breaks down (i.e. when samples of the unperturbed system poorly represent the perturbed system) is identified by divergent error estimates on the density perturbation. The polynomial degree is determined by a primitive type of dictionary learning~\cite{brunton2022data}, where the fidelity of the fit is monitored as the number of terms is reduced. When a substantial decrease is observed in the fitting performance, it becomes evident that too many terms have been removed. In this step, the fidelity metric is averaged over noise realisations added to the data to remove correlations between estimates at different $A$. When the structure of the polynomial has been determined, the final polynomial fitting procedure is wrapped in a Jackknife error estimate procedure~\cite{berg2004markov} to produce the estimate of the coefficients with well-founded error bars. The Jackknife is performed over independent simulations with different initial seeds of the random number generator. Further details of the fitting are provided in Appendix~\ref{sec:polynomial}.

\section{Density response from single harmonic perturbation}\label{sec:single_pert}

The PIMC results presented in this work have been obtained using the open-source \texttt{ISHTAR} code~\cite{ISHTAR}, which uses a canonical extended ensemble implementation~\cite{dornheim2021ab} of the worm algorithm by Boninsegni \etal~\cite{boninsegni2006worm_a,boninsegni2006worm_b}. Specifically, we consider the system of a spin-unpolarised uniform electron gas under periodic boundary conditions, with the degeneracy parameter $\Theta = 1$. Two different density conditions are examined: $r_s = 3.23$ and $r_s = 10.0$, which span the density that is possible to achieve in hydrogen jet experiments~\cite{zastrau2021high,fletcher2022electron,hamann2023prediction} to the boundary of the strongly coupled electron liquid regime, in which the dispersion relation has been predicted to be strongly modified~\cite{takada2016emergence,koskelo2023shortrange,dornheim2022electronic,dornheim2018ab}. The results for the two conditions will be shown as needed to highlight their differences.

\subsection{Density response of uniform electron gas}\label{sec:response_result}

\begin{figure*}
    \centering
    \includegraphics[width=\linewidth]{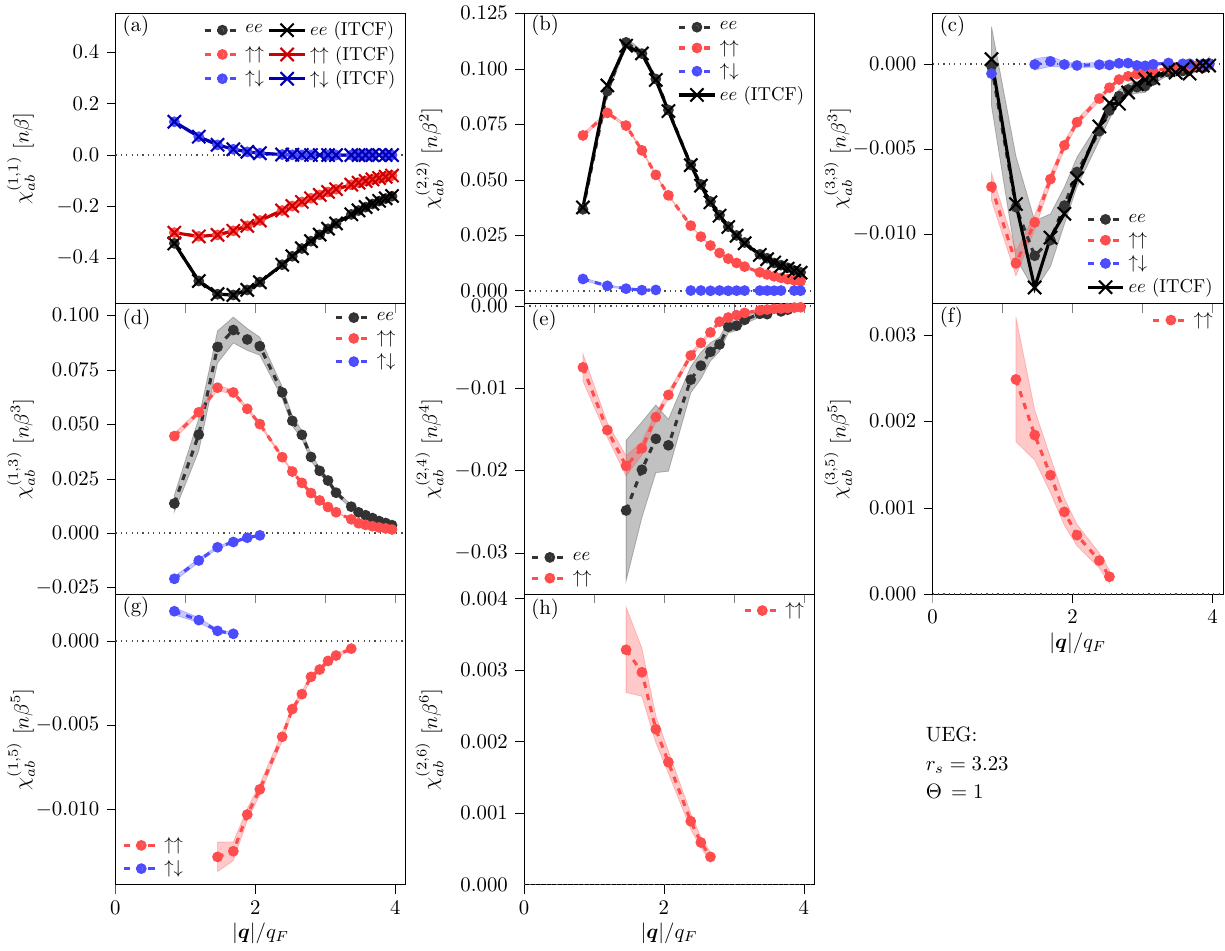}
    \caption{Density response coefficients for the UEG at $r_s = 3.23$ and $\Theta = 1.0$ with $N = 14$ and $P = 50$. The response coefficients are defined via the polynomial expansion of Eq.~\eqref{eq:response_singel_pert}. Results for the first harmonic (left column, three leading order coefficients), second harmonic (middle column, three leading order coefficients) and third harmonic (right column, two leading order coefficients) of the spin averaged ($ee$, black), spin diagonal ($\uparrow\uparrow$, red) and spin off-diagonal ($\uparrow\downarrow$, blue) response computed via the reweighting method (circles). The spin off-diagonal response decays rapidly for large wavenumbers, and the fitting procedure cannot reliably estimate the respective response coefficients. The shaded areas show $95\%$ confidence intervals estimated as described in Section \ref{sec:diag_response_estimators}. Results derived from the ITCF estimator~\cite{dornheim2021nonlinear,dornheim2022spin} are shown for comparison (crosses).}
    \label{fig:response_rs_323}
\end{figure*}

The results of the reweighting estimator approach, described in Section \ref{sec:diag_response_estimators}, are shown in Figure \ref{fig:response_rs_323} for $r_s = 3.23$. The leading order response at each harmonic is also estimated with the ITCF approach~\cite{dornheim2021nonlinear,dornheim2022spin}, the details of which are provided in Appendix~\ref{sec:ITCF}. The expected excellent agreement between the two approaches is confirmed for the first and second harmonics for the spin-averaged and spin-resolved electron responses, while the results at the third harmonic lie within the estimated statistical error. The results shown have been computed using $P = 50$ imaginary time slices and $N = 14$ electrons. In the present setup, the statistical uncertainty is lower in the ITCF method compared to the reweighting method. This holds even though the results were taken from two separate simulation runs using the same computational resources, where substantially fewer samples are generated with the ITCF estimators due to its comparatively costly evaluation of higher order responses. In fact, as discussed in Appendix \ref{sec:ITCF}, the quadratic and cubic ITCF estimators have a computational cost of $\order{P^3}$ and $\order{P^4}$, respectively. The reweighting estimator costs $\order{P}$ for each harmonic, and accordingly the reweighting method can be the preferred method when using a large number of imaginary time slices. Further comparisons between the two methods that are indicative of this trend are shown in Appendix~\ref{sec:statistical_error}. 

A larger relative uncertainty for higher harmonics and the sub-leading terms can be discerned in Figure \ref{fig:response_rs_323}. These terms are associated with higher powers of $A$ in the perturbation expansion and are smaller in the considered regime. Consequently, the higher order contributions are more challenging to measure above the statistical noise, and the reweighting method is generally helped by a more significant system response. Therefore, the response for larger $r_s$ (stronger coupling) requires less computational effort to converge, as the average sign increases, leading to less statistical noise~\cite{dornheim2015permutation}. Additional comparisons between coupling conditions are provided in Appendix~\ref{sec:additional}.

\begin{figure*}
    \centering
    \includegraphics[width=0.75\linewidth]{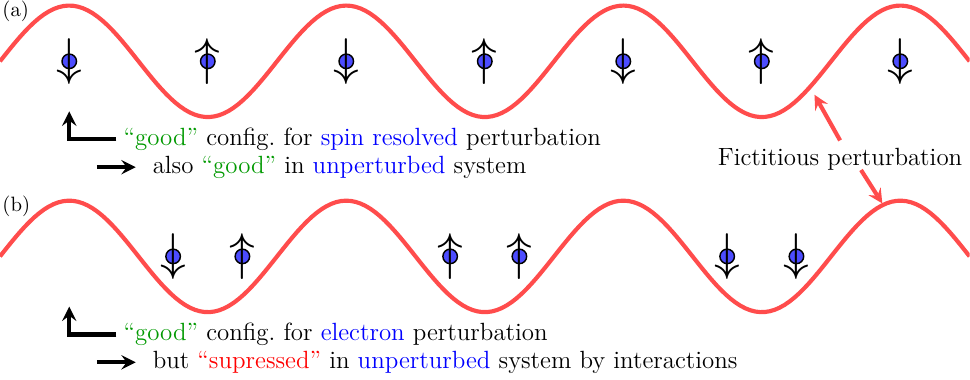}
    \caption{Schematic illustrating two idealised one-dimensional configurations which are ``good'' for measuring (a) spin resolved and (b) total electron response. ``Good'' in this instance refers to configurations with a larger statistical weight. Electrons are shown as circles along with an arrow indicating their spin. The fictitious perturbation is shown as the solid curve.}
    \label{fig:schematic}
\end{figure*}

As demonstrated in Figure \ref{fig:response_rs_323}, the reweighting estimator method gives access to species resolved response functions also for the higher order response. It turns out that the species diagonal response (e.g., $c = \uparrow$ and $\Bar{c} = \uparrow$) is easier to measure compared to the total electron response at all orders. This response is less challenging to measure as the associated free energy change between the unperturbed and perturbed systems is smaller when only a single spin species is perturbed and the reweighting factor is closer to unity. This allows for the use of larger $A$ in the fitting procedure, which is especially helpful in resolving higher order response coefficients, as evident from the lower rows of Figure \ref{fig:response_rs_323}. For a heuristic understanding, see the two idealised one-dimensional configurations in Figure \ref{fig:schematic}. Configuration (a) features well-spaced particles, a comparatively low inter-particle interaction energy, and therefore a large statistical weight in the unperturbed system. It also has a larger statistical weight when perturbing only the spin-up electrons that are located close to the potential wells of the fictitious perturbation. Consequently, this type of configuration is good for sampling in both unperturbed and perturbed systems, and the reweighting can thus be relatively well performing. On the other hand, consider configuration (b) which is a desirable configuration to sample in the system where all electrons are perturbed, as no electrons are close to the maxima of the fictitious perturbation. However, this configuration has smaller particle separations, and thus its statistical weight is suppressed in the unperturbed system. Configuration (b) will therefore not be sampled as often in unperturbed simulations resulting in a worse performing reweighting. The species resolved reweighting can thus be more efficient than the total response estimator, as the non-perturbed species can occupy the positions close to the fictitious potential maxima, resulting in configurations with lower interaction energy.

Estimation of the spin off-diagonal response (e.g., $c = \downarrow$ and $\Bar{c} = \uparrow$) is associated with the same free energy change as the spin diagonal response and the configurations that are desirable to sample are the same. However, the off-diagonal response is still more challenging to resolve. This response is mediated purely via particle interactions, as the external perturbations do not act on the measured species. Hence, especially at large $\vec{q}$ (single particle regime), the off-diagonal response is weak and more statistics are needed to resolve the density response above the statistical noise. The perturbation approaches are aided by a more substantial response.

\begin{figure*}
    \centering
    \includegraphics[width=\linewidth]{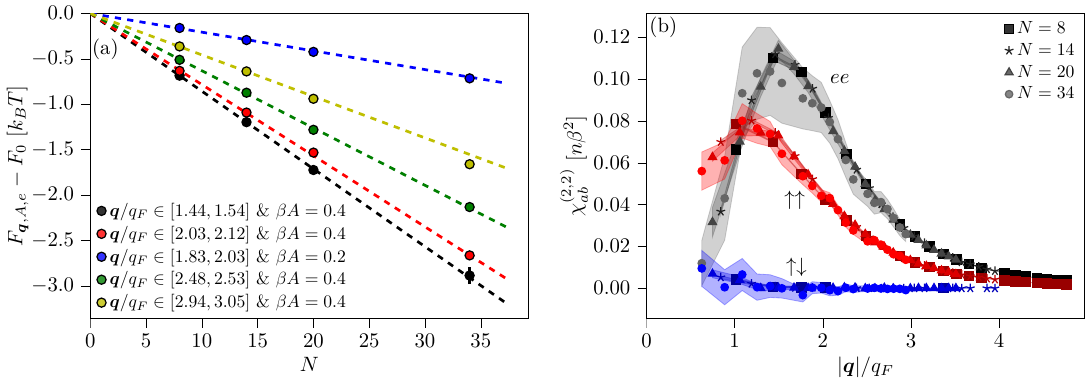}
    \caption{Results for the UEG at $r_s = 3.23$ and $\Theta = 1.0$ with $P = 50$. (a) Free energy change due to a harmonic perturbation, as computed from the reweighting estimator (circles) for varying perturbation wavenumbers and perturbation amplitudes detailed in the legend. Results are grouped based on similar wave numbers; identical wavenumbers are not accessible for varying system sizes due to the finite simulation box. 
    The dashed lines are linear fits with respect to the system size $N$. (b) Quadratic response at the second harmonic, as computed by the reweighting estimator for four numbers of electrons $N$. Results are shown for the spin averaged ($ee$, black), spin diagonal ($\uparrow\uparrow$, red) and spin off-diagonal ($\uparrow\downarrow$, blue) response. The shaded areas show $95\%$ confidence intervals. The cases $N = 14$ and $N = 20$ used the same amount of computational resources, while the cases $N = 8$ and $N = 34$ used $-85\%$ and $+160\%$ computational resources, respectively.}
    \label{fig:N_depdendnet_response}
\end{figure*}

The modeling of larger system sizes is generally desirable to reduce finite size errors. Fortunately, wavenumber resolved quantities, such as the density response, generally exhibit weaker finite size effects compared to bulk scale averaged properties~\cite{chiesa2006finte,dornheim2016ab,dornheim2021overcoming}. Still, modeling a larger number of particles $N$ is still desirable, since it reduces the smallest $\vec{q}$ that is possible to resolve in periodic simulations. In section \ref{sec:diag_response_estimators_estimator}, it was postulated that the change in free energy for the harmonically perturbed system would be proportional to $N$, which is indeed what is observed in Figure \ref{fig:N_depdendnet_response}(a). Results with minor variations in $\vec{q}$ are grouped due to the varying size of the simulation box, and deviations from the linear trend are attributed to this variation in $\vec{q}$. The linear scaling with $N$ will have a further theoretical explanation in Section \ref{sec:stiffnes_theorem}, but the result is that the reweighting factor in the denominator of Eq.~\eqref{eq:generla_reweighting} grows exponentially with $N$, and the reweighting method for a given $A$ is less efficient for larger system sizes. In the language of Figure \ref{fig:schematic}, as the number of particles increases, the unperturbed simulations are less likely to generate configurations where none of the particles are close to the maxima of the fictitious perturbation. 

The effect of the system size dependent free energy change is directly seen in Figure \ref{fig:N_depdendnet_response}(b) where $\chi^{(2,2)}_{ee}$ is estimated for varying number of particles while the same set of perturbation amplitudes $A$ has been used. The computational resources used in Figure \ref{fig:response_rs_323}, which led to well converged results for $N = 14$, are taken as the baseline. The results for the smaller system with $N = 8$ converge well with only $15\%$ of the computational cost, while the results for the larger system $N = 34$ still show substantial statistical variation using more than $2.5$ times the computational resources. A challenging computational scaling with $N$ is typical for the standard fermionic sign problem, but now the reweighting also becomes less efficient. This could be partially amended by considering smaller amplitudes $A$ with increasing $N$, but then the system response is smaller and more difficult to resolve over statistical noise. Having discussed the properties of the finite size errors, we now turn to consider the discretisation error due to a finite number of propagators.

\subsection{Propagator errors in harmonically perturbed systems}\label{sec:propagator_errors}

\begin{figure*}
    \centering
    \includegraphics[width=\linewidth]{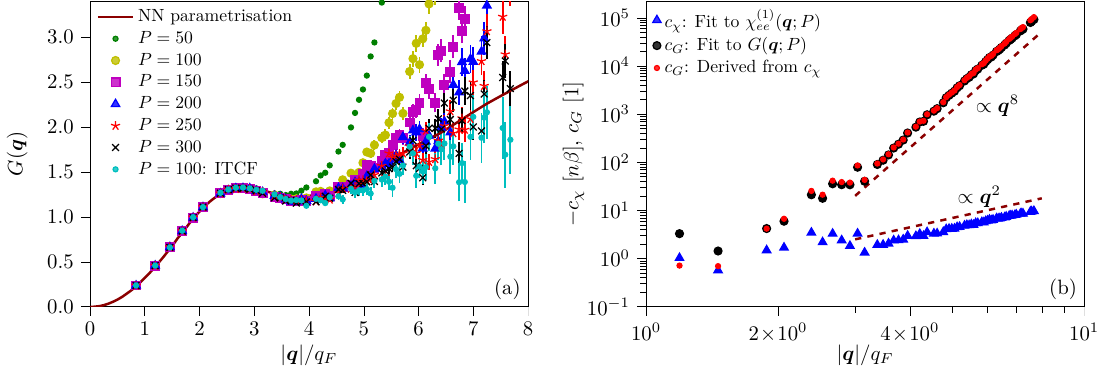}
    \caption{Finite propagator number ($P$) effects for the UEG at $r_s = 10$ and $\Theta = 1.0$ with $N = 14$. (a) Comparison of the extracted LFCs with $P=50,\,100,\,150,\,200,\,250,\,300$ showing that a large number of imaginary time slices are necessary for convergence at large $\vec{q}$. The error bars correspond to a standard deviation as computed by a Jackknife estimator. For comparison, extracted LFCs from the ITCF method with $P = 100$ and a neural network (NN) parametrisation~\cite{dornheim2019static} are also shown. (b) Coefficients for the $P^{-2}$-error in $\chi_{ee}^{(1)}(\vec{q})$ ($c_{\chi}(\vec{q})$, triangles) and $G(\vec{q})$ ($c_G(\vec{q})$, circles). In the latter case, results derived from $c_{\chi}(\vec{q})$ and Eq.~\eqref{eq:c_G} (small red circles), as well as direct fits to $G(\vec{q}; P)$ (big black circles) are shown. The quadratic and the octic dashed scaling lines are shown for reference.}
    \label{fig:P_dependence}
\end{figure*}

The PIMC formulation is exact within statistical error in the limit of infinite number of propagators $P$~\cite{ceperley1995path}. Therefore, convergence with respect to $P$ should be investigated for each quantity of interest. The (static) local field correction (LFC) can be viewed as a recasting of the (static) linear density response $\chi^{(1)}_{ee}(\vec{q})$ that highlights interaction effects beyond the mean-field level. It is defined as
\begin{equation}
    G(\vec{q}) = 1 - v_{\vec{q}}^{-1} \left( \frac{1}{\chi^{(1)}_{0}(\vec{q})} - \frac{1}{\chi^{(1)}_{ee}(\vec{q})} \right),
    \label{eq:LCF_definition}
\end{equation}
with $v_{\vec{q}}$ the Fourier transform of the Coulomb pair potential and $\chi^{(1)}_{0}$  the ideal linear density response function. Within the random phase approximation (RPA), $G(\vec{q})$ vanishes, and a non-zero $G(\vec{q})$ corresponds to higher order interaction effects. Therefore, differences are commonly more pronounced in the LFC than in the density response itself, especially for $|\vec{q}| \gg q_F$ where the density response is close to the ideal response. In Figure \ref{fig:P_dependence}(a), the discretisation error due to a finite $P$ is highlighted by comparing the LFC at $r_s = 10$ for $P$s between $50$ and $300$. Essentially, no differences can be discerned in the predictions for $|\vec{q}|$ below $3q_F$, and good agreement is observed with a previous neural network parameterisation that was trained on cases with $|\vec{q}| < 5 q_F$~\cite{dornheim2019static}. However, substantial differences are observed for the reweighting scheme for $|\vec{q}| > 4q_F$; at the largest $|\vec{q}| \approx 8 q_F$ the LFC might not have converged for $P = 300$. For large $|\vec{q}|$, the LFC should increase quadratically with a coefficient related to the excess kinetic energy~\cite{farid1993extremal,hou2022exchange}, but converging the asymptote using the reweighting estimator would be challenging. 

It is well known that the leading order factorisation error is proportional to $[[\Hat{V}, \Hat{K}], \Hat{V}] / P^2$~\cite{sakkos2009high}, where $\Hat{K}$ is the kinetic energy operator and $\Hat{V} = \Hat{V}_{0} + \Hat{V}_{\text{ext}}$ is the potential energy operator which is the sum of the inter-particle interaction energy $\Hat{V}_{0}$ and the external potential $\Hat{V}_{\text{ext}}$. Expanding the commutators for the case of the harmonically perturbed system, one arrives at
\begin{equation}
    \begin{aligned}
        [[\Hat{V}, \Hat{K}], \Hat{V}] = \frac{\hbar^2}{m} &\sum_{j = 1}^{N} \Big( 4A^2 \vec{q}^2 \sin^2(\vec{q}\cdot\Hat{\vec{x}}_j)\\
        + &4A  \vec{q}\cdot \Hat{\vec{F}}_j^{0} \sin(\vec{q}\cdot\Hat{\vec{x}}_j) + |\Hat{\vec{F}}_j^{0}|^2 \Big),
    \end{aligned}
    \label{eq:propagator_error}
\end{equation}
where $\Hat{\vec{x}}_j$ is the position operator of particle $j$ and $\Hat{\vec{F}}_j^{0}$ is the force operator from inter-particle interactions on the same particle~\cite{sakkos2009high}. Note that the expected factorisation error has the standard contribution $|\Hat{\vec{F}}_j^{0}|^2$, but also terms originating from the external perturbation. The proportionality factor for the factorisation error is extracted from a fit of the form $\chi^{(1)}_{ee}(\vec{q}; P) = \chi^{(1)}_{ee}(\vec{q}) + c_{\chi}(\vec{q}) P^{-2}$, where $\chi^{(1)}_{ee}(\vec{q}; P)$ is the estimated response using $P$ propagators. The resulting error coefficients are shown in Figure \ref{fig:P_dependence}(b). The $|\Hat{\vec{F}}_j^{0}|^2$ term will have a $\vec{q}$ independent contribution which could explain the close to constant error observed for small $\vec{q}$, but for large $\vec{q}$ a clear quadratic trend $c_{\chi} \propto \vec{q}^2$ is observed which is explained by the first term in Eq.~\eqref{eq:propagator_error}. 

The propagator error for the LFC in Figure \ref{fig:P_dependence}(a) has a stronger $\vec{q}$-dependence compared to $\chi_{ee}^{(1)}$. This is a result of the particular structure of Eq.~\eqref{eq:LCF_definition}. If the propagator error $c_{\chi}(\vec{q}) P^{-2}$ in $\chi_{ee}^{(1)}$ is propagated to the LFC, the resulting error model is $G(\vec{q}; P) = G(\vec{q}) + c_G(\vec{q}) P^{-2} + \order{P^{-4}}$, where $G(\vec{q}; P)$ is the LFC evaluated for a given number of propagators $P$ and
\begin{equation}
    c_G(\vec{q}) = - \frac{v_{\vec{q}}^{-1} c_{\chi}(\vec{q})}{\left(\chi_{ee}^{(1)}(\vec{q}) \right)^2}.
    \label{eq:c_G}
\end{equation}
The large $\vec{q}$ behaviour of linear density response functions is dictated by $\chi_{ee}^{(1)}(\vec{q}) \approx \chi_{0}^{(1)}(\vec{q}) \propto \vec{q}^{-2}$. Therefore, in this limit, the dominant scaling is $c_{G}(\vec{q}) \propto \vec{q}^{6} c_{\chi}(\vec{q}) \propto \vec{q}^8$. The predicted octic scaling for the LFC error is observed to match the results of direct fits to $G(\vec{q}; P)$ which are included in Figure \ref{fig:P_dependence}(b). In general, the reweighting estimator approach is not ideal for studying the fine details of the density response at large $\vec{q}$, but this is not a characteristic of the reweighting method \textit{per se}; it is believed to be a general feature of direct perturbation approaches within the primitive factorisation. 

Potential avenues to address the $P$ error would be to include the external potential in higher order factorisation schemes along the lines of Chin~\cite{chin2015high} or the external potential could be directly incorporated into the single particle propagator using \textit{Mathieu functions} as the basis instead of plane waves. The latter option has some structural similarities to previous work which incorporates external harmonic oscillator potentials in the single particle propagator~\cite{weiss2005path,karmakar2026combining}. Both of these methods have the potential to remove bias and improve convergence with respect to the number of imaginary time slices but would require some reformulation of the reweighting factor.

As a final remark, let us contrast the observed $P$ error behaviour to the corresponding error in the ITCF based method. In this alternative method, no external perturbation is applied to the system, so $A = 0$ in Eq.~\eqref{eq:propagator_error}, and generally a small $P$-error is observed in the ITCF data~\cite{groth2019ab}. However, for the density response estimation, the ITCF is integrated over imaginary time with a discretisation $\epsilon = \beta / P$, and the assosiated discretisation error can be the dominant finite $P$ error (see the additional discussion in Appendix~\ref{sec:ITCF}). However, in our tests using a fifth order integrator in Figure \ref{fig:P_dependence}(a), the ITCF method using $P = 100$ performed similarly to the perturbation method using $P = 300$. Therefore, the ITCF method currently remains the preferred method for investigating the large $\vec{k}$ limit. 

\subsection{Nonlinear density stiffness theorem}\label{sec:stiffnes_theorem}

The energy and free energy changes that accompany the application of a harmonic perturbation to a quantum system have attracted substantial theoretical interest in the ground state and at finite temperature~\cite{moldabekov2018theoretical,dornheim2023energy,moldabekov2025density}, respectively. Their relation to static density response properties is known as the \textit{density stiffness theorem}~\cite{giuliani2008quantum}. In the ground state, this relation has proven useful for the evaluation of density response coefficients from computational methods where energy is a convenient observable~\cite{moroni1992static,moroni1995static}, results that have been utilised for the development of exchange correlation functionals~\cite{perdew2008restoring,tao2008nonempirical,sun2015strongly}. The difference in free energy between two systems in PIMC can be evaluated by thermodynamic integration~\cite{frenkel2002understanding} or by the newly introduced $\eta$-ensemble~\cite{dornheim2025direct,dornheim2025eta}, which recently demonstrated the ability to reach a large number of particles~\cite{dornheim2025fermionic,svensson2025accelerated}. In the reweighting estimator approach, the free energy difference is directly related to the denominator of the estimators and is directly measured.

Gravel and Ashcroft~\cite{gravel2007nonlinear} used thermodynamic integration to calculate the change in the free energy in terms of density response coefficients to an arbitrary order in the external perturbation. Their results can be straightforwardly generalised to multi-component systems. For a uniform system, the change in (total) free energy is
\begin{widetext}
\begin{equation}
    F_{b} - F_{a} = \sum_{n = 1}^{\infty} \frac{1}{n+1} \sum_{\Bar{c}} \sum_{\vec{k}_1, c_1} \frac{V^{c_1}_{\text{ext}}(\vec{k}_1)}{\Omega} \cdots \sum_{\vec{k}_n, c_n} \frac{V^{c_n}_{\text{ext}}(\vec{k}_n)}{\Omega}\, V^{\Bar{c}}_{\text{ext}}(-\vec{k}_1 - \cdots - \vec{k}_n)\, \chi^{(n)}_{\Bar{c}c_1\cdots c_n}(\vec{k}_1, \dots, \vec{k}_n),
    \label{eq:density_stiffnes_general}
\end{equation}
\end{widetext}
where $V_{\text{ext}}^{c_i}(\vec{k})$ is the Fourier transform of $V_{\text{ext}}^{c_i}(\vec{r})$. In the above, the sums ($c_i$, $\Bar{c}$) are over the species in the system and over the $\vec{k}$-vectors allowed in a finite simulation box with periodic boundary conditions. If the perturbation is specialised to the case in Section \ref{sec:diag_response_estimators} with a single harmonic perturbation acting on one species, Eq.~\eqref{eq:density_stiffnes_general} simplifies to
\begin{equation}
     \frac{F_{\vec{q}, A, c} - F_{0}}{N} = n^{-1} \sum_{m = 1}^{\infty} \frac{\chi_{cc}^{(1,2m-1)}(\vec{q})}{m} A^{2m},
     \label{eq:density_stiffnes}
\end{equation}
where $n = N / \Omega$ is the density. This provides a nonlinear generalisation to the linear density stiffness theorem that is valid at finite temperature.

Consider the structure of Eq.~\eqref{eq:density_stiffnes}. First, the free energy change is extensive and proportional to the number of particles, as previously shown in Figure \ref{fig:N_depdendnet_response}(a). Second, only the species diagonal response and density perturbations in phase with the original perturbation contribute to the free energy change, i.e. the first harmonic response. Methods that rely on the free energy difference to extract response properties~\cite{moroni1992static,moroni1995static} are thus more limited in the available response properties. The presented estimators can therefore provide additional responses, which could act as additional constraints for the development of exchange correlation functionals. Last, as a corollary to the second point, only terms even in $A$ contribute, as the response at the first harmonic is odd.

\begin{figure}
    \centering
    \includegraphics[width=\linewidth]{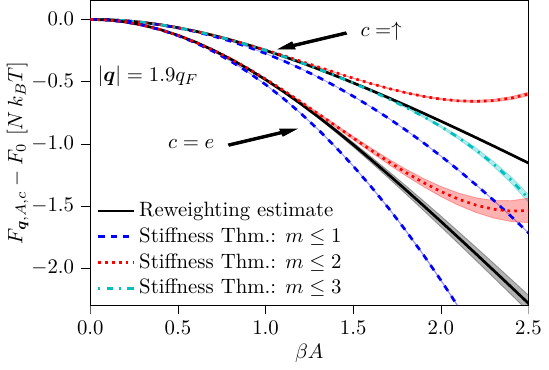}
    \caption{Free energy change due to a harmonic perturbation, as computed from the reweighting estimator (solid black) for the UEG at $r_s = 3.23$ and $\Theta = 1.0$ with $N = 14$ and $P = 50$. The perturbation wavenumber is $|\vec{q}|/q_F=1.9$ and the perturbation amplitude varies. Results are shown for the perturbations of all electrons ($e$, bottom) and a single spin orientation ($\uparrow$, top), as indicated in the figure. Results are compared to the predictions of the nonlinear density stiffness theorem with one (dashed blue), two (dotted red), or three (dashed-dotted cyan) terms included in the series expansion with respect to the perturbation order, see the right hand side of Eq.~\eqref{eq:density_stiffnes}. The density response coefficients are adopted from Figure \ref{fig:response_rs_323}. The shaded areas are estimated $95\%$ confidence intervals.}
    \label{fig:stiffnes_theorem}
\end{figure}

In Figure \ref{fig:stiffnes_theorem}, the direct evaluation of the free energy difference from the reweighting estimator is compared to the nonlinear density stiffness theorem. For small $A$, the agreement between the two calculations is so good that we need to emphasise that Figure \ref{fig:stiffnes_theorem} does not comprise a fit, but that the response coefficients used in Eq.~\eqref{eq:density_stiffnes} are those shown in Figure \ref{fig:response_rs_323}. As $A$ progressively increases, additional terms in Eq.~\eqref{eq:density_stiffnes} become significant and require the estimation of higher order response properties. This explains the smaller discrepancy for the spin resolved perturbation at the largest $A$ values, since higher order coefficients have been possible to estimate for the spin resolved case. The excellent agreement with the density stiffness theorem provides strong numerical evidence for Eq.~\eqref{eq:density_stiffnes} in addition to the theoretical derivation. At this point, it should be emphasised that the nonlinear density stiffness theorem provided in Ref.~\cite{dornheim2023energy} is not correct.

\section{Density response estimation with additional species resolution}\label{sec:mult_pert_species}

The estimators presented up to this point involve perturbing a single species ($c$) and measuring the response on the same or another species ($\Bar{c}$). As seen in Eq.~\eqref{eq:coefficent_definition_single_perterbation}, this structure of the estimator gives one access to the $n$th order response function with one species index being $\Bar{c}$ and the remaining $n$ species indices being $c$. Therefore, for quadratic and higher order response functions, the previously described estimator cannot access all cross-species contributions. The remaining cross-species response can be explored by fictitiously perturbing multiple species simultaneously.

\subsection{Density response estimators for additional species resolution}\label{sec:species_estimator}

Let us consider harmonic perturbations acting on two species $c_1$ and $c_2$, i.e., the external potential is
\begin{equation}
    \Hat{V}_{\text{ext}} = 2A \sum_{i = 1}^2 \int\! d\vec{r}\; \cos(\vec{q} \cdot \vec{r}) \Hat{n}^{c_i}(\vec{r}),
    \label{eq:perterbation_two_species}
\end{equation}
where the amplitudes and wavenumbers of the perturbations are the same for simplicity. The perturbed system is now labelled $b \equiv (\vec{q}, A, c_1, c_2)$. This perturbed system $b$, can be investigated using the reweighting formulation, if the sum over particles in the reweighting factor \eqref{eq:reweighting_factor} is taken to be over all particles of species $c_1$ and $c_2$.

If the system response is measured at $\vec{q}$ and higher order harmonics, the response has the same structure as Eq.~ \eqref{eq:response_singel_pert}, but the response coefficients now have an additional species index, e.g., $\chi^{(n,m)}_{\Bar{c}c_1c_2}$. The linear response at the first harmonic becomes the superposition of responses from perturbing each species individually:
\begin{subequations}
\begin{equation}
    \chi^{(1,1)}_{\Bar{c}c_1c_2}(\vec{q}) = \chi^{(1)}_{\Bar{c}c_1}(\vec{q}) + \chi^{(1)}_{\Bar{c}c_2}(\vec{q}).
\end{equation}
These response functions could already be probed when perturbing a single species. However, at the second harmonic, the leading-order response is
\begin{equation}
    \begin{aligned}
        \chi^{(2,2)}_{\Bar{c}c_1c_2}(\vec{q}) =\, &\chi^{(2)}_{\Bar{c}c_1c_1}(\vec{q},\vec{q}) + 2\chi^{(2)}_{\Bar{c}c_1c_2}(\vec{q},\vec{q})\\
        +\, &\chi^{(2)}_{\Bar{c}c_2c_2}(\vec{q},\vec{q}),
    \end{aligned}
\end{equation}
including the previously inaccessible cross-species density response $\chi^{(2)}_{\Bar{c}c_1c_2}(\vec{q},\vec{q})$. If $\chi^{(2,2)}_{\Bar{c}c_1c_2}$ is combined with $\chi^{(2,2)}_{\Bar{c}c_1}$, $\chi^{(2,2)}_{\Bar{c}c_2}$ from our previous estimators, all species combinations for the quadratic response function can be isolated. At the third harmonic, the leading-order response is
\begin{equation}
    \begin{aligned}
        \chi^{(3,3)}_{\Bar{c}c_1c_2}(\vec{q}) = &\chi^{(3)}_{\Bar{c}c_1c_1c_1}(\vec{q},\vec{q},\vec{q}) + 3 \chi^{(3)}_{\Bar{c}c_1c_1c_2}(\vec{q},\vec{q},\vec{q})\\
        + &\chi^{(3)}_{\Bar{c}c_2c_2c_2}(\vec{q},\vec{q},\vec{q}) + 3\chi^{(3)}_{\Bar{c}c_1c_2c_2}(\vec{q},\vec{q},\vec{q}),
    \end{aligned}
    \label{eq:species_third_order_third_harmonic}
\end{equation}
and the cubic coefficient at the first harmonic is
\begin{equation}
    \begin{aligned}
        \frac{\chi^{(1,3)}_{\Bar{c}c_1c_2}(\vec{q})}{3}\! = &\Bar{\chi}^{(3)}_{\Bar{c}c_1c_1c_1}(\shortminus\vec{q},\vec{q},\vec{q}) +\! 3 \Bar{\chi}^{(3)}_{\Bar{c}c_1c_1c_2}(\shortminus\vec{q},\vec{q},\vec{q})\\
        + &\Bar{\chi}^{(3)}_{\Bar{c}c_2c_2c_2}(\shortminus\vec{q},\vec{q},\vec{q}) +\! 3\Bar{\chi}^{(3)}_{\Bar{c}c_1c_2c_2}(\shortminus\vec{q},\vec{q},\vec{q}).
    \end{aligned}
    \label{eq:species_third_order_first_harmonic}
\end{equation}
\end{subequations}
For the cubic response, the response function has four species indices and three species should be perturbed, for complete access to all combinations of species. However, in the special case of two-component systems, the present formulation is enough to explore all species combinations due to the permutation symmetry in the species index. In the following section, this estimator will be demonstrated for the UEG.

\subsection{Nonlinear cross-species response for the uniform electron gas}\label{sec:species_results}

\begin{figure}
    \centering
    \includegraphics[width=\linewidth]{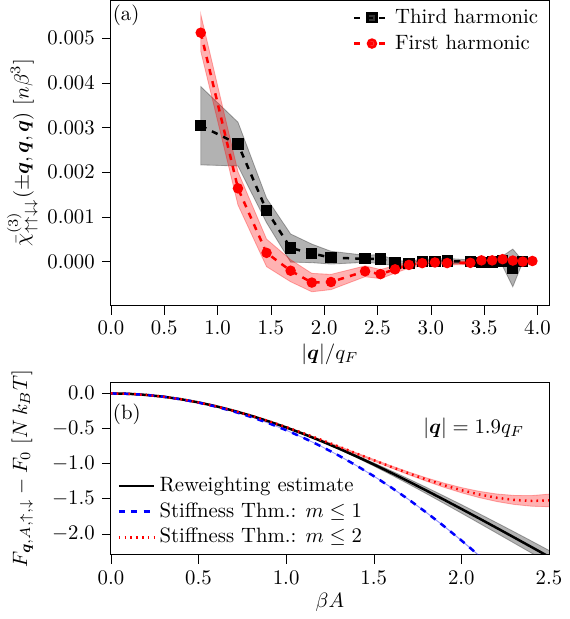}
    \caption{The nonlinear cross-species response of the UEG at $r_s = 3.23$ and $\Theta = 1.0$ with $N = 14$ and $P = 50$. In both plots, the shaded areas are estimated $95\%$ confidence intervals. (a) Cubic density response coefficients for the species combination not accessible when perturbing a single species (see Figure \ref{fig:response_rs_323}). Results for the third (black squares) and first (red circles) harmonic responses. (b) Free energy change due to a simultaneous harmonic perturbation in two species. The perturbation wavenumber is $|\vec{q}|/q_F=1.9$ and the perturbation amplitude varies. Reweighting estimator results (solid black) are compared to the predictions of the nonlinear density stiffness theorem with one (dashed blue) and two (dotted red) terms included in the series expansion with respect to the perturbation order, see the right hand side of Eq.\eqref{eq:density_stiffnes_species}. The response coefficients are derived from the same dataset as (a).}
    \label{fig:species_estimator}
\end{figure}

The unpolarised uniform electron gas is a special case of a two-component system since there is complete symmetry between the two spin orientations. Therefore, the four unique species combinations for the quadratic response function can be grouped into two groups that can be explored with the original estimators of Section \ref{sec:diag_response_estimators}. This will not be the case for more general two-component systems such as the electron-ion plasma or the partially spin-polarised UEG. However, aiming to exemplify the additional response functions accessible using the estimator in Section \ref{sec:species_estimator}, the cubic response will be investigated.

Using equations \eqref{eq:species_third_order_third_harmonic} and \eqref{eq:species_third_order_first_harmonic}, the response functions $\chi_{\uparrow\uparrow\downarrow\downarrow}^{(3)}(\vec{q}, \vec{q}, \vec{q})$ (third harmonic) and $\Bar{\chi}_{\uparrow\uparrow\downarrow\downarrow}^{(3)}(-\vec{q}, \vec{q}, \vec{q})$ (first harmonic) can be estimated, respectively. Figure \ref{fig:species_estimator}(a) shows results for these cubic response functions that are not accessible when perturbing a single species. Their general structure is characterized by a rapid decay for large $|\vec{q}|$. However, the sign change in the cubic response at the first harmonic for intermittent $|\vec{q}|$ should be pointed out. At the smallest $|\vec{q}|$ resolved, $\chi_{\uparrow\uparrow\downarrow\downarrow}^{(3)}(\vec{q}, \vec{q}, \vec{q})$ is of similar magnitude to $\chi_{\uparrow\downarrow\downarrow\downarrow}^{(3)}(\vec{q}, \vec{q}, \vec{q})$ but of the opposite sign.

The general version of the nonlinear density stiffness theorem, Eq.~\eqref{eq:density_stiffnes_general}, can be adapted to cases where two species are perturbed simultaneously. The result is similar to Eq.~\eqref{eq:density_stiffnes}, but as the perturbation couples to two species, the response of both species at the first harmonic contributes to the free energy change, i.e.,
\begin{equation}
    \begin{aligned}
        \frac{F_{\vec{q}, A, c_1, c_2} - F_{0}}{N} = n^{-1} \sum_{m = 1}^{\infty} \frac{1}{m} \Big( &\chi_{c_1 c_1 c_2}^{(1,2m-1)}(\vec{q})\\ 
        + &\chi_{c_1 c_2 c_2}^{(1,2m-1)}(\vec{q}) \Big) A^{2m}.
    \end{aligned}
     \label{eq:density_stiffnes_species}
\end{equation}
A comparison between the estimated free energy change and Eq.~\eqref{eq:density_stiffnes_species}, is shown in Figure \ref{fig:species_estimator}(b). The results show excellent agreement for small perturbation amplitudes $A$, but the discrepancy increases for large $A$ as additional coefficients need to be estimated.  

\section{Complete quadratic density response}\label{sec:mult_pert_q}

Until now, the response due to perturbations at a single harmonic has been considered, providing information on nonlinear response functions along certain cuts in the wavenumber space. For the evaluation of nonlinear corrections to properties such as effective ion interactions, a complete description of all $\vec{k}$-vector combinations is required~\cite{brovman1974phonons,nagao2003enhanced,gravel2007nonlinear,porter2010pair}. To explore the complete quadratic response function, we will now consider the reweighting methodology with two fictitious harmonic potentials that stimulate the system at two different wavevectors.

\subsection{Complete quadratic density response estimators}\label{sec:complete_response_estimators}

The external perturbation of interest is
\begin{equation}
    \Hat{V}_{\text{ext}} = 2A\sum_{i = 1}^{2} \int\! d\vec{r}\; \cos(\vec{q}_i \cdot \vec{r}) \Hat{n}^{c}(\vec{r}),
    \label{eq:two_perterbation}
\end{equation}
where for simplicity only one species $c$ is perturbed and the amplitudes of the two perturbations are assumed to be equal. The structure of the response in such systems was explicitly provided by Dornheim \etal~\cite{dornheim2022nonlinear}. Here we focus particularly on the response at the wavenumber $\vec{k} = \vec{q}_1 + \vec{q}_2$:
\begin{widetext}
\begin{equation}
    \left\langle \Hat{\rho}^{\Bar{c}}_{\vec{q}_1 + \vec{q}_2} \right\rangle_{b} = 
    \begin{cases}
        \hspace{0.6cm}\hphantom{\chi^{(1)}_{\Bar{c}c}(\vec{q}_i) A + {}} 4 \chi^{(2)}_{\Bar{c}cc}(\vec{q}_1, \vec{q}_2) A^2 + \order{A^4} & \text{if } \vec{q}_1 = \vec{q}_2,\\
        \hspace{4.45cm} 0 & \text{else if } \vec{q}_1 + \vec{q}_2 = 0,\\
        \hspace{0.6cm}\chi^{(1)}_{\Bar{c}c}(\vec{q}_i) A + 2 \chi_{\Bar{c}cc}^{(2)}(\vec{q}_1, \vec{q}_2) A^2 + \order{A^3} & \text{else if } \vec{q}_1 + \vec{q}_2 = -\vec{q}_i\hphantom{2} \text{ for } i = 1, 2,\\
        \left[\chi_{\Bar{c}cc}^{(2)}(\vec{q}_i, \vec{q}_i) + 2\chi_{\Bar{c}cc}^{(2)}(\vec{q}_1, \vec{q}_2) \right] A^2 + \order{A^4} & \text{else if } \vec{q}_1 + \vec{q}_2 = - 2\vec{q}_i \text{ for } i = 1, 2,\\
        \hspace{0.6cm}\hphantom{\chi^{(1)}_{\Bar{c}c}(\vec{q}_i) A + {}} 2 \chi_{\Bar{c}cc}^{(2)}(\vec{q}_1, \vec{q}_2) A^2 + \order{A^3} & \text{else if } \vec{q}_1 + \vec{q}_2 = \pm 3\vec{q}_i \text{ for } i = 1, 2,\\
        \hspace{0.6cm}\hphantom{\chi^{(1)}_{\Bar{c}c}(\vec{q}_i) A + {}} 2 \chi_{\Bar{c}cc}^{(2)}(\vec{q}_1, \vec{q}_2) A^2 + \order{A^4} & \text{else},
    \end{cases}
    \label{eq:quadratic_response_two_perterbations}
\end{equation}
\end{widetext}
where it has been assumed that $\vec{q}_1 \neq 0$ and $\vec{q}_2 \neq 0$. The case $\vec{q}_1 + \vec{q}_2 = 0$ vanishes due to the fixed number of particles in canonical simulations, which can be seen as a shift of the chemical potential~\cite{moldabekov2025generalized}. As the two perturbation vectors can be chosen independently, the complete quadratic response function can be explored, except for the special case of $\vec{q}_1 + \vec{q}_2 = 0$. It is pointed out that when $\vec{q}_1 + \vec{q}_2$ is equal to the negative of one of the perturbation vectors, the quadratic response is not the leading order response, and both even and odd $A$ powers emerge in the perturbative expansion. This will ultimately result in larger statistical errors for this case and has required a minor modification of the polynomial fitting procedure (see Appendix~\ref{sec:polynomial}). A similar scenario occurs when $\vec{q}_1 + \vec{q}_2$ coincides with the third harmonic, but in this case the quadratic response is of the leading order. Furthermore, when $\vec{q}_1 + \vec{q}_2$ is equal to twice the negative of one perturbation vector, the desired response function appears in combination with the response function on the diagonal. Fortunately, this response function can be extracted from the estimator in Section \ref{sec:diag_response_estimators} and subtracted from the result. As the reweighting factor can readily be modified to the case with two perturbations and the fitting procedure can incorporate the structure of Eq.~\eqref{eq:quadratic_response_two_perterbations}, a complete quadratic response reweighting estimator is achieved.

Dornheim \etal~\cite{dornheim2022nonlinear} provided an estimator for the quadratic response function resolved over two $\vec{k}$-vectors, based on three-body imaginary time correlation functions. They applied this estimator to a few select combinations of $\vec{q}_1$ and $\vec{q}_2$, and successfully benchmarked the results against computations with direct perturbations. In Appendix~\ref{sec:ITCF}, a summary of the method is given. We have re-implemented their estimators in the \texttt{ISHTAR} code to be able to consider a larger number of combinations for $\vec{q}_1$ and $\vec{q}_2$, and combined it with a Jackknife estimator for the error analysis. This ITCF estimator will serve as a comparison for the reweighting estimator in the following.

\begin{figure*}
    \centering
    \includegraphics[width=\linewidth]{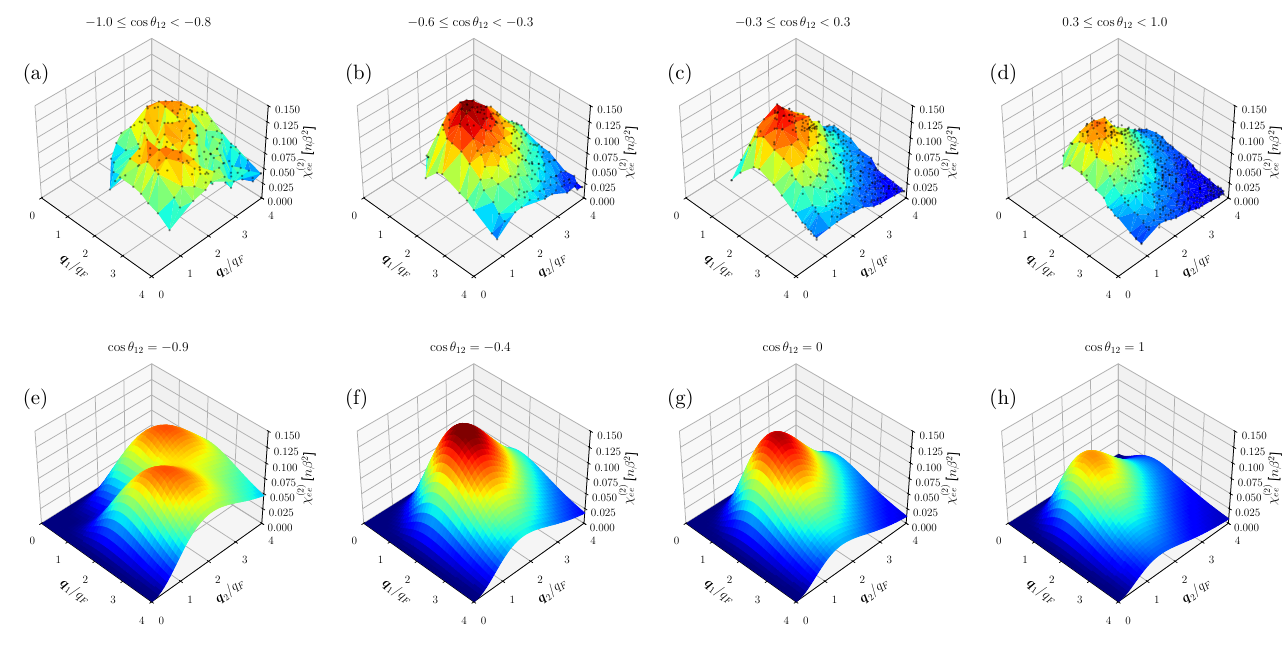}
    \caption{The complete quadratic response of the UEG at $r_s = 3.23$ and $\Theta = 1.0$. The upper row (a) -- (d) depicts PIMC results obtained with the reweighting estimator method with $P = 50$ and $N = 14$. Each sub-figure groups results for different angles $\theta_{12}$ between $\vec{q}_1$ and $\vec{q}_2$, which are specified by the top label. Each dot corresponds to a single estimate. The lower row (e) -- (h) depicts results obtained with the approximation of Eq.~\eqref{eq:quadratic_response_approx}, where the LFC has been computed by the reweighting estimator method and has been interpolated. In contrast to the top row, results are provided for specific angles $\theta_{12}$.}
    \label{fig:complete_quadratic_response}
\end{figure*}

\subsection{Complete quadratic density response for uniform electron gas}\label{sec:complete_response_result}

The complete quadratic response function $\chi_{ee}^{(2)}(\vec{k}_1,\vec{k}_2)$ in a uniform system is a function of two wave vectors, and therefore the input space is six dimensional. Although, for isotropic and homogeneous systems, the quadratic response function depends only on the magnitude of the two $\vec{k}$-vectors ($|\vec{k}_1|$ and $|\vec{k}_2|$) and the angle between such $\theta_{12}$~\footnote{The presented simulations are performed in a cubic simulation cell which formally breaks isotropy. However, as finite size effects were shown to be small in Figure \ref{fig:N_depdendnet_response}, we believe this still to be a good approximation.}. These symmetries allow us to substantially reduce the number of combinations considered.

In Figures \ref{fig:complete_quadratic_response}(a)--(d), the complete (static) quadratic response function is shown. To our knowledge, this is the first time the complete two dimensional spectrum of the quadratic response function has been shown from \textit{ab initio} computations. The corresponding results have not been derived from the direct perturbation method, as the number of $\vec{k}$-vector combinations needed to be considered grow prohibitively large. However, using the reweighting scheme, all of the 634 $\vec{k}$-vector combinations in Figure \ref{fig:complete_quadratic_response}(a)--(d) have been derived from a single set of simulations of the unperturbed system. For $\cos\theta_{12} > -0.6$, the quadratic response function is observed to have a single peak at $\vec{q}_1 \approx \vec{q}_2 \approx 2q_F$, while when $\vec{q}_1$ and $\vec{q}_2$ are pointing almost in the opposite direction, $\cos\theta_{12} \lesssim -0.8$, this peak is split into two. To further reason about the results shown in Figure \ref{fig:complete_quadratic_response}(a)-(d), consider Eq.~\eqref{eq:quadratic_response_two_perterbations} where the system responds to harmonic perturbations at $\vec{q}_1$ and $\vec{q}_2$, and the response is measured at $\vec{q}_1 + \vec{q}_2$. For perturbations with small wavelengths, the linear response of a quantum particle vanishes due to a finite de Broglie wavelength~\cite{dornheim2024ab}, something we also observe in the quadratic case. The above is the most evident in the forward direction $\cos\theta_{12} > 0$, where the ``measurement wave number'' $|\vec{q}_1 + \vec{q}_2|$ is larger than the individual perturbation wave numbers, and the response decays rapidly. Conversely, the response when $\cos\theta_{12} < 0$ can be larger than in the forward direction, since the measurement wave number is not as large. The response in the limit of small wave numbers can be quantitatively understood on RPA level. Large scale density perturbations are suppressed due to screening, and the response at small $\vec{q}_1$ or $\vec{q}_2$ is thus suppressed. In the far backward direction $\cos\theta_{12} \lesssim -0.8$, the measurement wave number becomes small and a response here would correspond to a large length scale perturbation and is thus suppressed by screening. This results in the aforementioned double peak featured in the response function.

\begin{figure}
    \centering
    \includegraphics[width=\linewidth]{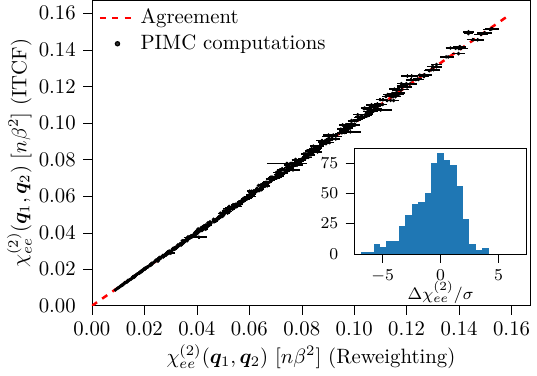}
    \caption{Comparison between the complete quadratic response of the UEG from the reweighting estimator approach and the ITCF approach at $r_s = 3.23$ and $\Theta = 1.0$ with $N = 14$ and $P = 50$. The error bars show estimated $95\%$ confidence intervals. Very close agreement is observed between the methods. The inset shows the difference between the results, normalised with the uncertainty $\sigma = \sqrt{\sigma_{\text{RW}}^2 + \sigma_{\text{ITCF}}^2}$, where $\sigma_{\text{RW}}$ and $\sigma_{\text{ITCF}}$ are the estimated standard deviations for the reweighting estimator approach and the ITCF approach, respectively.}
    \label{fig:comp_complte_quadratic}
\end{figure}

The PIMC results are further compared with model responses in Figures \ref{fig:complete_quadratic_response}(e)--(h). Previous investigations of the quadratic density response suggested that the quadratic response function can be approximated by~\cite{paasch1980quadratic,dornheim2021density,moldabekov2025generalized}
\begin{equation}
    \chi_{ee}^{(2)}(\vec{k}_1, \vec{k}_2) \approx \frac{\chi_{0}^{(2)}(\vec{k}_1, \vec{k}_2)}{\varepsilon(\vec{k}_1) \varepsilon(\vec{k}_2) \varepsilon(\vec{k}_1+\vec{k}_2)},
    \label{eq:quadratic_response_approx}
\end{equation}
where $\varepsilon(\vec{k}) = 1 - v_{\vec{k}}[1 - G(\vec{k})]\chi^{(1)}_0(\vec{k})$. When $G(\vec{k}) = 0$ this corresponds to the well-known RPA approximation for quadratic response functions~\cite{paasch1977quadratic,hu1988z,pitarke1995quadratic,vorberger2025green}. Including a linear order correction in terms of local field corrections has been found to be effective for $\vec{k}_1 = \vec{k}_2$~\cite{dornheim2021density}. In equation \eqref{eq:quadratic_response_approx}, $\chi_{0}^{(2)}(\vec{k}_1, \vec{k}_2)$ is the ideal (static) quadratic response function. For the diagonal case $\vec{k}_1 = \vec{k}_2$, the recursion relation given by Mikhailov~\cite{mikhailov2012second,mikhailov2014nonlinear} and generalised by Tolias \etal~\cite{tolias2023unravelling}, can be used for its evaluation at any temperature. However, to our knowledge, the complete description of $\chi^{(2)}_{0}$ is only available at zero temperature~\cite{lloyd1968structural,milchev1977quadratic,hu1988z,wang1992kinetic,pitarke1995quadratic,rommel1998quadratic,bergara1999quadratic}. For the current evaluation, the treatment of Cenni \etal~\cite{cenni1988evaluation,cenni1992fermionic} has been generalised to finite temperature by substituting the appropriate distribution function and numerically performing the integration over the momentum magnitude.

From the comparison between the PIMC results and Eq.~\eqref{eq:quadratic_response_approx} in Figure \ref{fig:complete_quadratic_response} it is observed that the approximation that only uses linear order corrections captures the main features of the quadratic response function in the complete $\vec{k}$-vector space. The agreement is also quantitatively good everywhere except for the smallest $\vec{k}$-vectors where higher order corrections are seen to be more substantial. Hence, when evaluating nonlinear corrections to properties that strongly depend on the long-range screening behaviour, more sophisticated approximations than Eq.~\eqref{eq:quadratic_response_approx} will be required. Descriptions of the complete quadratic response function have previously been limited to the local density approximation (LDA) in the ground state~\cite{nagao2003enhanced,porter2010pair}, and have never been benchmarked in the full space. Consequently, the current data set opens up the possibility of constructing and testing more sophisticated approximations for the complete quadratic response function, information which could further be used to constrain exchange correlation functions.

The reweighting estimator results for the quadratic response functions are tested against the ITCF method results in Figure \ref{fig:comp_complte_quadratic}. An excellent agreement is found between the two methods for all $\vec{k}$-vector combinations considered. In general, we observe that the statistical error tends to be larger for the reweighting method compared to the ITCF method for the simulations with $N = 14$ and $P = 50$, which is in general agreement with the results of Section \ref{sec:response_result} and Appendix \ref{sec:statistical_error}. Thus, in cases with rather small $P$, the ITCF method should be still preferred. However, due to the rapidly increasing computational cost of the quadratic response ITCF estimator with respect to $P$, the reweighting estimator will be a more efficient method in the case of a large $P$ number.

\section{Conclusions}\label{sec:conclusions}

Static density response properties are central for the understanding of quantum systems including warm dense matter, with a direct relation to exchange-correlation functionals, scattering cross-sections and stopping powers. Path integral Monte Carlo methods enable high fidelity simulations of quantum systems, but in order to evaluate density response properties one either performs simulations with an explicit external potential or considers imaginary time correlation functions. However, some Monte Carlo methods do not provide access to imaginary time correlation data (e.g., restricted path integral Monte Carlo~\cite{ceperley1991fermion}), and the direct perturbation approach generally requires a large number of simulations for similar systems. Furthermore, the ITCF approach~\cite{dornheim2021ab} is not yet available for many offdiagonal response functions, such as the cubic response at the first harmonic~\cite{vorberger2025green}. To overcome this, we introduced a new density response estimator based on a reweighting procedure, where samples from the unperturbed system are reweighted so that averages in multiple perturbed systems can be evaluated in parallel. This allows us to estimate the full spectrum of static density response properties purely from samples of the unperturbed system without using imaginary time relations. The reweighting estimator can potentially be used in restricted PIMC simulations, although the impact of the virtual perturbation on the nodal surfaces, and hence the corresponding Monte Carlo configuration space, might require special considerations.

It is demonstrated that the introduced reweighting estimator converges for the uniform electron gas at moderate degeneracy. The properties of the reweighting estimator with respect to the particle number and imaginary time slices have also been analysed. In particular, it is observed that the reweighting method is more suited to investigate smaller size systems, as with increasing system size, the free energy difference between the perturbed and unperturbed systems increases and the efficiency of the reweighted samples decreases. Furthermore, it is demonstrated that the finite propagator number error for the density response increases quadratically with the wave number $\vec{q}$ for large $|\vec{q}|$. However, this bias is believed to exist for all direct perturbation methods and not to be intrinsic to the reweighting scheme.

Having explored the properties that can be extracted from a single harmonic perturbation, the methodology is readily adapted to scenarios with multiple simultaneous perturbations. With two fictitious perturbations possessing different wave-vectors, mode coupling between external perturbations acting on different species can be used to derive additional nonlinear species off-diagonal density response functions as well as the full quadratic response function resolved over two vector arguments. The complete quadratic response will become particularly valuable for testing new theoretical models for the quadratic density response functions. The two analysed extensions to the reweighting estimator could be combined, and potentially extra perturbations could be considered. Thus, the method has substantial flexibility to explore almost any desired nonlinear static density response function.  

Furthermore, we note that the present approach is in no way limited to path integral Monte Carlo or the modeling of UEG and WDM systems, and can be readily used in future works to study the density response of a gamut of quantum many-body systems such as ultracold atoms~\cite{filinov2010berezinskii,ferre2016dynamic,dornheim2022path} and even material mixtures~\cite{boninsegni1995path,boninsegni2017kinetic}. Finally, the present approach is formulated for any observable. Thus, it can be used for the microscopic current density vector operator instead of the microscopic density operator. This would unveil the structure of the nonlinear current density response tensors, which are connected to current density functional theory~\cite{giuliani2008quantum}. In that case, the perturbed quantity is the vector potential.

\begin{acknowledgments}

This project has received funding from the Fusion2024 program of the German Federal Ministry of Research, Technology and Space (BMFTR) via the project "VANLIFE" (funding no 13F1016B), and the European Research Council (ERC) under European Union’s Horizon 2022 research and innovation programme (Grant agreement No.101076233, "PREXTREME"). Views and opinions expressed are however those of the authors only and do not necessarily reflect those of the European Union or the European Research Council Executive Agency. Neither the European Union nor the granting authority can be held responsible for them. This work has received funding from the German Federal Ministry of Research, Technology and Space (BMFTR) via the ErUM Data project "DEMOS" (05D25CR1). This work has received funding from the European Union's Just Transition Fund (JTF) within the project \emph{R\"ontgenlaser-Optimierung der Laserfusion} (ROLF), contract number 5086999001, co-financed by the Saxon state government out of the State budget approved by the Saxon State Parliament.
Tobias Dornheim gratefully acknowledges funding from the Deutsche Forschungsgemeinschaft (DFG) via project DO 2670/1-1.
Computations were performed on a Bull Cluster at the Center for Information Services and High-Performance Computing (ZIH) at Technische Universit\"at Dresden and the Norddeutscher Verbund f\"ur Hoch- und H\"ochstleistungsrechnen (HLRN) under grant mvp00024.
\end{acknowledgments}

\appendix

\section{Polynomial fitting procedure}\label{sec:polynomial}

The extraction of the harmonic density response coefficients with the direct perturbation and the reweighting estimator methods involves polynomial fitting. Especially when considering multiple different harmonic perturbations or multiple species combinations, as possible with the reweighting method, a more systematic approach to the polynomial fitting problem is required. The approach followed in this manuscript was summarised in Section \ref{sec:diag_response_estimators}, and further details are provided below.

\begin{figure*}
    \centering
    \includegraphics[width=0.49\linewidth]{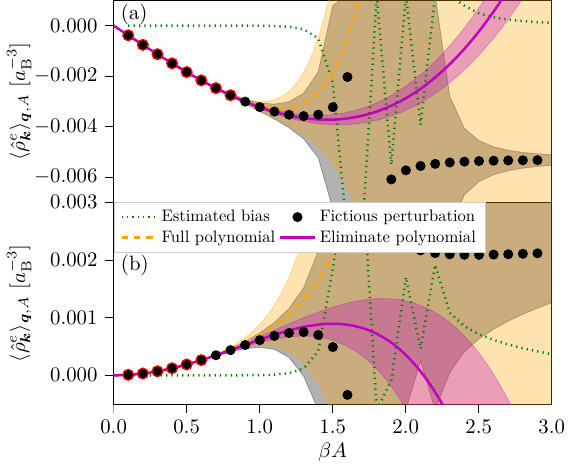}
    \includegraphics[width=0.49\linewidth]{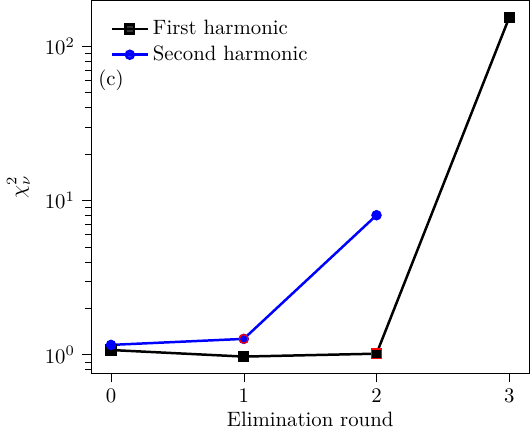}
    \caption{Polynomial fitting procedure for harmonic perturbations acting on all electrons with $|\vec{q}| = 1.46\,q_F$, for the UEG at $r_s = 3.23$ and $\Theta = 1.0$ with $N = 14$ and $P = 50$. Examples of fitting using the full polynomial (dashed orange) in Eq.~\eqref{eq:response_singel_pert} limited to a highest order of $7$, and using the elimination procedure (solid magenta) are shown for the first and second harmonics in (a) and (b), respectively. The shaded areas are estimated $95\%$ confidence intervals. The points used for the fit are highlighted with an additional circle and a bias estimate is shown (dotted green)~\cite{berg2004markov}. (c) Reduced $\chi^2$ in the elimination process to determine the number of polynomial coefficients for the first harmonic (black squares, case a) and second harmonic (blue circles, case b), respectively. An additional border highlights the selected polynomial based on the condition \eqref{eq:jump_cond} in each case.}
    \label{fig:polynomial_fitting}
\end{figure*}

The reweighting estimator is evaluated for a set of $A$ values for each wavenumber, but for larger perturbation strengths the reweighting factor $\langle \Hat{1}_{ab} \rangle_{a}$ increases. The result is a worse performing estimator with increasing statistical error. For a large enough $A$, the system $b$ is no longer ergodically explored by the reweighted samples of system $a$ giving rise to a divergent error estimate $\delta \langle \Hat{\rho}^{\Bar{c}}_{\vec{q}}\rangle$ and a large estimated bias, here evaluated by a Jackknife estimator~\cite{berg2004markov}. The breakdown of the reweighting procedure has been observed before, when the sampled and targeted distributions are too different~\cite{ceperley1979quantum}. These characteristics are exemplified in Figures \ref{fig:polynomial_fitting}(a) and (b). Therefore, in what follows, only values of $A$ below this point are considered. We fit from $A \in [0, A_{\text{max}}]$, where the maximum perturbation strength $A_{\text{max}}$ is such that 
\begin{equation}
    \delta \langle \Hat{\rho}^{\Bar{c}}_{\vec{k}} \rangle_{\vec{q}, A, c} \leq \varepsilon \langle \langle \Hat{\rho}^{\Bar{c}}_{\vec{k}} \rangle_{\vec{q}, A, c}\rangle_A
\end{equation}
for all $A \leq A_{\text{max}}$. For the results reported in this work, $\varepsilon$ in the range of $10^{-3} - 2\times 10^{-1}$ was used.

The bounded domain $[0, A_{\text{max}}]$ indicates that density perturbations can be fit by a finite order polynomial. The general structure of the perturbation expansion has the restricted form based on Eq.~\eqref{eq:response_singel_pert} and then using a dictionary learning (DL) algorithm, we can identify the smallest order necessary to represent the data. Our DL algorithm uses a dimensionality scan, commonly seen in introductory data science courses~\cite{brunton2022data}. Specifically, we optimise the monomial basis $\vec{n}$ based on the cost function
\begin{equation}
    \chi^2_{\nu} = \frac{ \left\langle \min_{\vec{c}} \left|\left| \frac{\mat{P}(\vec{n}, A_{\text{max}})\vec{c} - \langle \Hat{\rho}_{\vec{k}}^{\Bar{c}} \rangle_{\vec{q},A,c} + \vec{\eta} }{\delta \langle \Hat{\rho}_{\vec{k}}^{\Bar{c}} \rangle_{\vec{q},A,c}} \right|\right|^2 \right\rangle_{\vec{\eta}} }{N_A - ||\vec{n}||_1},
    \label{eq:cost_func}
\end{equation}
where $N_A$ is the number of $A$ points, $||\vec{n}||_{1}$ is the number of monomials, $\vec{c}$ is the polynomial coefficients, $\mat{P}(\vec{n},A_{\text{max}})$ is the design matrix for the monomial basis and the $A$ domain. The DL is performed by sequentially eliminating the highest order polynomial term. If $\chi^2_{\nu}$ remains approximately the same between elimination rounds, the eliminated monomial was not essential in describing the data. If a jump in $\chi^2_{\nu}$ is observed, the last eliminated term is essential to represent the data. Therefore, we select the polynomial that satisfies
\begin{equation}
    \chi^2_{\nu}(||\vec{n}||_1 - 1) \geq (1 + T) \chi^2_{\nu}(||\vec{n}||_1),
    \label{eq:jump_cond}
\end{equation}
where the threshold $T = 3$ has been used herein. 

One complication in the reweighting data set is that the different $A$ points cannot be considered statistically independent, as they are derived from the same samples. Therefore, when evaluating $\chi^2_{\nu}$ a Gaussian noise term $\vec{\eta}$ with standard deviation $\delta \langle \Hat{\rho}^{\Bar{c}}_{\vec{k}} \rangle_{\vec{q}, A, c}$ (estimated by a Jackknife over independent simulations) and mean zero is added to the data, and the cost function is averaged over $100$ such noise realisations ($\langle \cdot \rangle_{\vec{\eta}}$ in Eq.~\eqref{eq:cost_func}) to achieve a stable estimate. The elimination procedure is shown for two selected cases in Figure \ref{fig:polynomial_fitting}(c). Example fits are shown in Figures \ref{fig:polynomial_fitting}(a) and (b), where the better agreement outside of the fitting range after using the elimination procedure compared to using the largest polynomial is indicative of a well behaved fitting procedure.

In two cases for the complete quadratic density response estimator in Section \ref{sec:complete_response_estimators}, the fitting polynomial has both even and odd powers of $A$ in the perturbative expansion. With more terms, the polynomials have more flexibility to compensate for deficiencies in the fitting model, and the changes in $\chi_{\nu}^2$ between rounds are smaller. To guarantee a good fit in all cases for the complete quadratic response estimator, we also find the last round such that
\begin{equation}
    \chi_\nu^2 (||\vec{n}||_1) \leq 1 + \varepsilon_{\chi^2},
    \label{eq:val_cond}
\end{equation}
where $\varepsilon_{\chi^2} = 10^{-1}$, and select the larger of the two polynomials suggested by Eqs.~\eqref{eq:jump_cond} and \eqref{eq:val_cond}.

Having established the range of amplitudes and structure of the polynomial to use, a final estimate and associated errors are evaluated using a Jackknife estimator~\cite{berg2004markov}. That is, the density fluctuation is estimated based on the data from all independent simulations (seed) except one, the appropriate polynomial is fitted, and the harmonic density response coefficients are extracted. This procedure is repeated for all seeds, and the resulting coefficient distributions are used for the final estimate and error quantification. For the case of LFCs and quantities that combine multiple response properties, their computation is performed in the Jackknife, and the average is taken as the last step.

\section{ITCF density response estimators}\label{sec:ITCF}

Throughout this work, the newly introduced density response estimator is compared to different ITCF estimators for linear and nonlinear response properties~\cite{dornheim2023electronic}. For completeness, the methods will be summarised here, which will also allow us to look at the sources of error and computational complexity of the approach.

\begin{figure*}
    \centering
    \includegraphics[width=\linewidth]{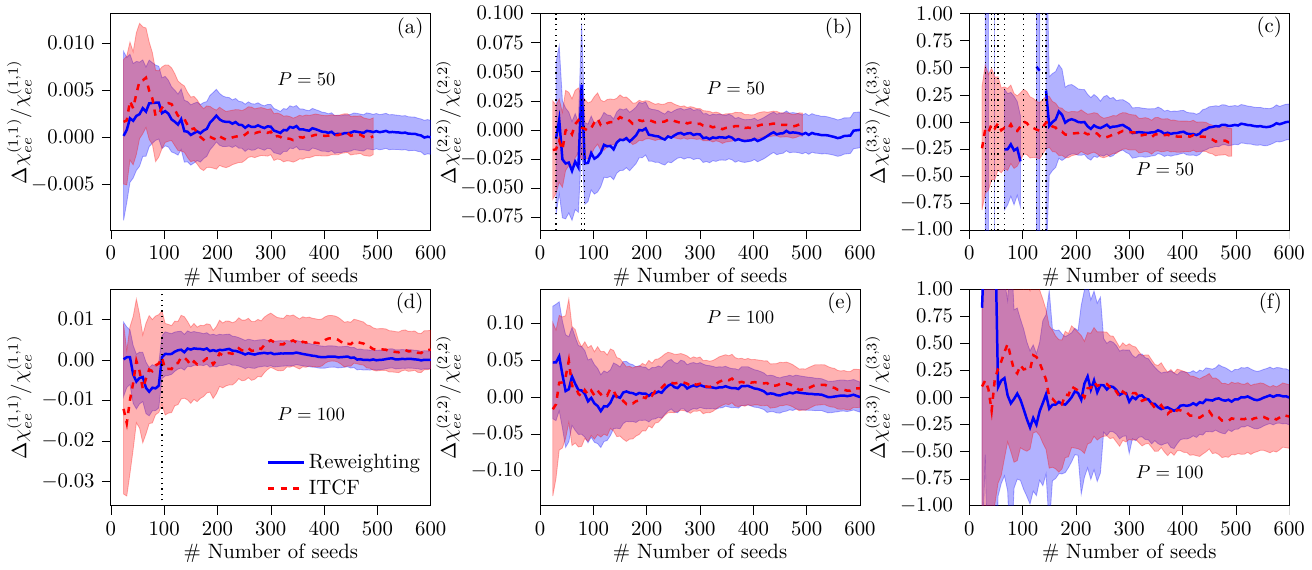}
    \caption{Estimates for $\chi^{(n,n)}_{ee}$ at $|\vec{q}| = 1.46q_F$ for the UEG at $r_s = 10$ and $\Theta = 1.0$ with $N = 14$, employing the reweighting (solid blue) and ITCF estimators (dashed red) with a varying number of independent simulations used for the estimation. Results are shown as the difference from the reweighting estimator results using all the available data and are normalised to the same value. The upper and lower rows show the cases $P = 50$ and $P = 100$, respectively. The left, middle and right columns illustrate the leading response at different orders $n = 1, 2, 3$, respectively. The shaded areas are estimated $95\%$ confidence intervals based on a Jackknife estimator. The dotted vertical lines highlight the points where the number of terms in the polynomials fits changes, as determined by the procedure described in Appendix \ref{sec:polynomial}.}
    \label{fig:error_comparison}
\end{figure*}

For linear response coefficients, the estimator takes the form of a simple integral over the imaginary time~\cite{groth2019ab,dornheim2021nonlinear,dornheim2022spin}
\begin{equation}
    \chi_{\Bar{c}c}^{(1)}(\vec{k}) = - n \int_{0}^{\beta}\! d\tau\; \frac{1}{N}\left\langle \Hat{n}_{\vec{k}}^{\Bar{c}}(\tau) \Hat{n}_{-\vec{k}}^{c}(0)  \right\rangle_0,
    \label{eq:linear_ITCF_estimator}
\end{equation}
where 
\begin{equation}
    \Hat{n}_{\vec{k}}^{c}(\tau) = \sum_{j = 1}^{N_c} e^{-i \vec{k} \cdot \Hat{\vec{r}}_{j}(\tau)},
\end{equation}
with $\Hat{\vec{r}}_{j}(\tau)$ the position operator of the particle $j$ at imaginary time $\tau$ and the sum over all particles of species $c$. Within the primitive factorisation in PIMC, the imaginary time $\tau$ is accessible on a discrete set of time slices $\tau_l = l \times \beta/P$ for $l = 0, \dots, P-1$. For the integration in Eq.~\eqref{eq:linear_ITCF_estimator}, we have employed a fifth order Newton–Cotes quadrature~\cite{abramowitz1948handbook}. Any quadrature will result in a ``discretisations'' error which can be viewed as a finite $P$ error, on top of the error in the integrand due to a finite number of imaginary time slices. The former can even in many scenarios be the dominant source of error. 

For the evaluation of $\Hat{n}_{\vec{k}}^{c}(\tau)$, $\order{N}$ computational work is required in the position basis, and this is evaluated for $P$ values of $\tau$ for each $\vec{k}$ vector. The resulting computational cost is therefore $\order{NP}$. The correlation function in the integral is evaluated for $P$ separations in imaginary time, but as there is cyclic translation invariance in imaginary time, we also average over the ``reference time slice'' where $\tau = 0$. Including this average result in a computational cost of $\order{P^2}$. Even though the average over reference time slice is not strictly needed, this is part of why the ITCF estimators converge quickly. The final numerical integration in Eq.~\eqref{eq:linear_ITCF_estimator} is a $\order{P}$ operation.

The structure of the ITCF estimator for the quadratic response is similar to the linear response case and takes the form~\cite{dornheim2021nonlinear,dornheim2022nonlinear}
\begin{equation}
    \begin{aligned}
        \chi^{(2)}_{cc}(\vec{k}_1, \vec{k}_2&) = \frac{n}{2} \int_0^{\beta}\!\! \int_0^{\beta}\!d\tau_1 d\tau_2\;\\
        &\frac{1}{N}\left\langle \Hat{n}_{\vec{k}_1+\vec{k}_2}^{c}(0) \Hat{n}_{-\vec{k}_1}^{c}(-\tau_1) \Hat{n}_{-\vec{k}_2}^{c}(-\tau_2)  \right\rangle_0.
    \end{aligned}
    \label{eq:quadratic_ITCF_estimator}
\end{equation}
The estimator for the second harmonic contribution is obtained when $\vec{k}_1 = \vec{k}_2$, but for the complete estimator $\vec{k}_1$ and $\vec{k}_2$ are generally different. The integration is now over two imaginary time arguments, and we use the fifth order Newton–Cotes integration recursively. The correlation function is evaluated for $P^2$ imaginary time arguments, and together with the averaging over the reference slice, the computational cost scales as $\order{P^3}$.

Finally, the cubic response ITCF estimator is~\footnote{Here we highlight a sign error in Eqs.~(21) and (23) of Ref.~\cite{dornheim2021nonlinear}.}
\begin{equation}
    \begin{aligned}
        \chi^{(3)}_{ccc}(\vec{k}&, \vec{k}, \vec{k}) = -\frac{n}{6} \int_0^{\beta}\!\! \int_0^{\beta}\!\! \int_0^{\beta}\!d\tau_1 d\tau_2 d\tau_3\;\\
        &\frac{1}{N}\left\langle \Hat{n}_{3\vec{k}}^{c}(0) \Hat{n}_{-\vec{k}}^{c}(-\tau_1) \Hat{n}_{-\vec{k}}^{c}(-\tau_2) \Hat{n}_{-\vec{k}}^{c}(-\tau_3)  \right\rangle_0,
    \end{aligned}
    \label{eq:cubic_ITCF_estimator}
\end{equation}
which now is a three dimensional integration. The computational cost of evaluating the correlation function is $\order{P^4}$ in analogy with arguments for lower order response functions. At this order, the scaling with $P$ can be substantially punishing, and only every $20$th imaginary time slice has been considered a reference slice when performing the average.  

\begin{figure*}
    \centering
    \includegraphics[width=\linewidth]{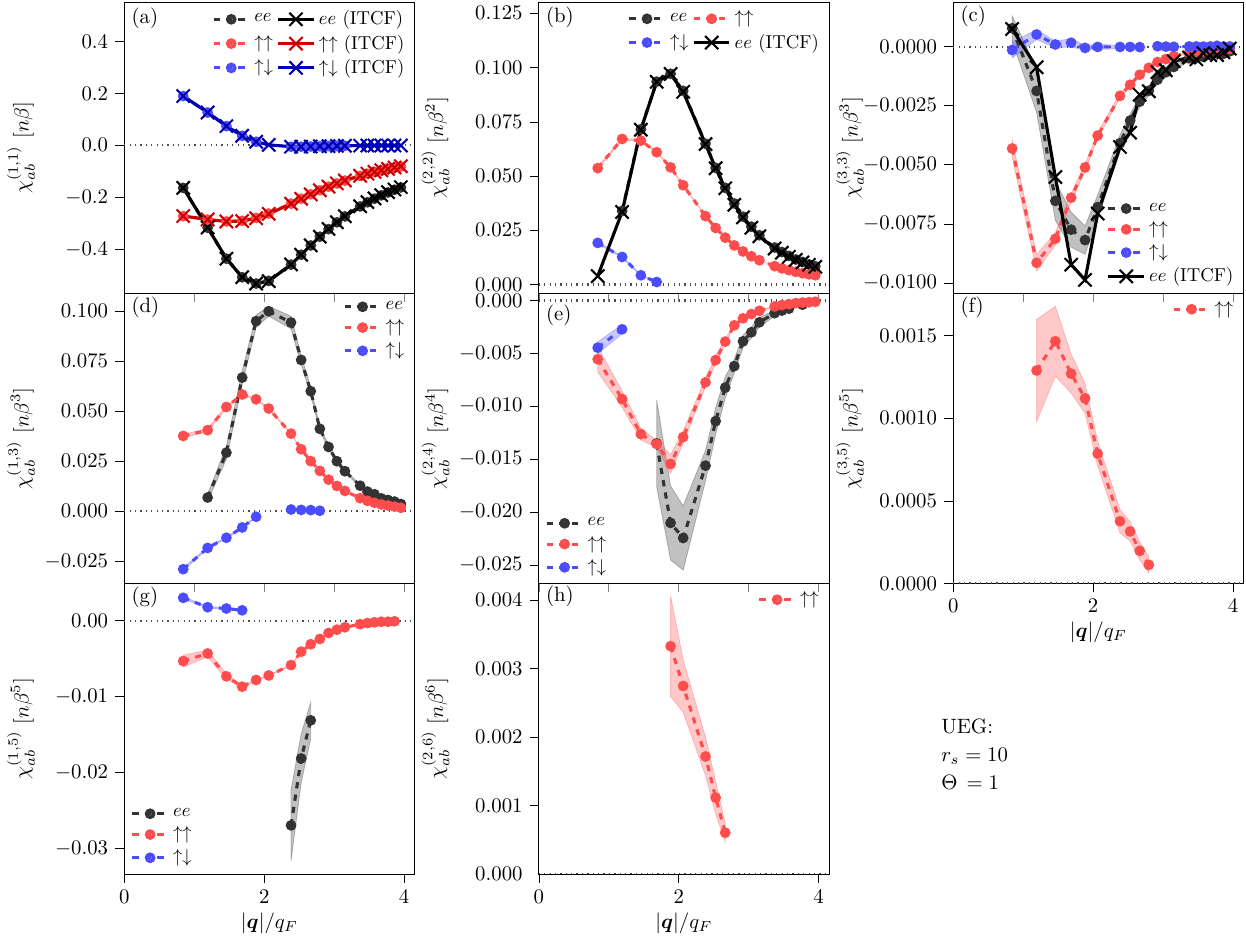}
    \caption{Density response coefficients for the UEG at $r_s = 10$ and $\Theta = 1.0$ with $N = 14$ and with $P = 50$. The response coefficients are defined via the polynomial expansion of Eq.~\eqref{eq:response_singel_pert}. Results for the first harmonic (left column, three leading order coefficients), second harmonic (middle column, three leading order coefficients) and third harmonic (two leading order coefficients) of the spin averaged ($ee$, black), spin diagonal ($\uparrow\uparrow$, red) and spin off-diagonal ($\uparrow\downarrow$, blue) response computed via the reweighting method (circles). The spin off-diagonal response decays rapidly for large wavenumbers, and the fitting procedure cannot reliably estimate the respective response coefficients. The shaded areas show $95\%$ confidence intervals estimated as described in Section \ref{sec:diag_response_estimators}. Results derived from the ITCF estimator~\cite{dornheim2021nonlinear,dornheim2022spin} are shown for comparison (crosses). The corresponding results for the UEG at $r_s = 3.23$ and $\Theta = 1.0$ are shown in Figure \ref{fig:response_rs_323}.}
    \label{fig:response_rs_10}
\end{figure*}

\section{Comparison of statistical error in the density response estimators}\label{sec:statistical_error}

In some applications, the presented reweighting estimator method and the ITCF method can access the same properties. However, when determining which method to use, the convergence rate of the methods in this particular case is of interest.

Figure \ref{fig:error_comparison} features a comparison of the estimated density response using the reweighting and ITCF methods for $r_s = 10$. In particular, two cases with $P = 50$ and $P = 100$ are depicted. In the $P = 50$ case of fewer imaginary time slices, the ITCF method has a smaller statistical error for the nonlinear density response, while the performance is similar for the linear response. The computations have been carried out expending the same computational resources, which results in fewer samples in the simulations using the ITCF method compared to the reweighting method. The higher order ITCF estimators are comparatively costly to evaluate, as the computational work scales as $\order{P^{n+1}}$ for the $n$th order estimator. Therefore, the smaller or similar statistical error achieved with the ITCF methods means that the intrinsic variance in the ITCF estimator is smaller than that in the reweighting estimator. Note that all the results shown in Figure \ref{fig:error_comparison} are taken from simulations where the cubic response and the complete quadratic response (Section \ref{sec:complete_response_result}) are estimated, and that convergence can be reached substantially quicker if we focus on estimating only one of the lower order response functions.

In the $P = 100$ case of more imaginary time slices, see Figure \ref{fig:error_comparison}, it is observed that the reweighting method has a smaller statistical variance for the linear and quadratic response. For cubic response, the estimated errors are roughly the same. This is a consequence of the rapidly increasing computational cost of the higher order ITCF estimator. The fidelity of the reweighting estimator is observed to decrease at larger harmonics; this is not due to an increased computational cost but rather due to the fact that the estimated response is smaller and conversely more challenging to extract from the polynomial fit. Nevertheless, the reweighting method can be the desired method for large $P$ simulations and higher order response estimators. 

\section{Density response results for $r_s = 10$}\label{sec:additional}

When increasing $r_s$, the Coulomb coupling in the UEG increases, which in turn increases the average sign and thus reduces the sign problem in the PIMC simulations. As a consequence, generally less computational effort is required for convergence~\cite{dornheim2015permutation}. A similar trend is observed for the reweighting estimator method, where for larger $r_s$ higher order harmonic coefficients can be estimated with less computational cost. In particular, the denominator $\langle \Hat{1}_{ab} \rangle_a$ in the reweighting estimator converges faster for larger $r_s$, even though its value is similar. This suggests that the faster convergence observed is related to the weaker fermionic PIMC sign problem.

For example, Figure \ref{fig:response_rs_10} features the estimated response coefficients for the UEG at $r_s = 10$ with $30\%$ less computational effort compared to the $r_s = 3.23$ results depicted in Figure \ref{fig:response_rs_323}. Yet, more higher order response coefficients can be reliably estimated at the first and second harmonic. The reweighting results for $r_s = 10$, generally follow the same conclusions as given in Section \ref{sec:response_result} for $r_s = 3.23$, and good agreement with the ITCF method is observed. This demonstrated the capability of reweighting methods to model the density response in the electron liquid regime.

\bibliography{ref}

@inbook{abramowitz1948handbook,
  title={Handbook of mathematical functions with formulas, graphs, and mathematical tables},
  author={Abramowitz, Milton and Stegun, Irene A},
  volume={55},
  year={1972},
  chapter={25.4},
  edition={10th},
  publisher={US Government printing office},
  address={Washington}
}

@article{baczewski2016x,
  title={X-ray {Thomson} scattering in warm dense matter without the {Chihara} decomposition},
  author={Baczewski, Andrew David and Shulenburger, L and Desjarlais, M P and Hansen, S B and Magyar, R J},
  journal={Physical Review Letters},
  volume={116},
  number={11},
  pages={115004},
  year={2016},
  doi={10.1103/PhysRevLett.116.115004},
  publisher={APS}
}

@book{berg2004markov,
  title={Markov chain {Monte} {Carlo} simulations and their statistical analysis: {With} web-based {Fortran} code},
  author={Berg, Bernd Albert},
  year={2004},
  publisher={World Scientific Publishing Company},
  address={Singapore}
}

@article{betti2016inertial,
  title={Inertial-confinement fusion with lasers},
  author={Betti, R and Hurricane, O A},
  journal={Nature Physics},
  volume={12},
  number={5},
  pages={435--448},
  year={2016},
  doi={10.1038/nphys3736},
  publisher={Nature Publishing Group}
}

@article{bergara1999quadratic,
  title={Quadratic electronic response of a two-dimensional electron gas},
  author={Bergara, A and Pitarke, J M and Echenique, P M},
  journal={Physical Review B},
  volume={59},
  number={15},
  pages={10145},
  year={1999},
  publisher={APS}
}

@article{bohme2022static,
  title={Static electronic density response of warm dense hydrogen: {Ab} initio path integral {Monte} {Carlo} simulations},
  author={B{\"o}hme, Maximilian and Moldabekov, Zhandos A and Vorberger, Jan and Dornheim, Tobias},
  journal={Physical Review Letters},
  volume={129},
  number={6},
  pages={066402},
  year={2022},
  publisher={APS}
}

@article{boninsegni1995path,
  title = {Path Integral Monte Carlo Simulation of Isotopic Liquid Helium Mixtures},
  author = {Boninsegni, Massimo and Ceperley, David M.},
  journal = {Physical Review Letters},
  volume = {74},
  issue = {12},
  pages = {2288--2291},
  numpages = {0},
  year = {1995},
  month = {Mar},
  publisher = {American Physical Society},
  doi = {10.1103/PhysRevLett.74.2288},
  url = {https://link.aps.org/doi/10.1103/PhysRevLett.74.2288}
}

@article{boninsegni2006worm_a,
author = {M. Boninsegni and N. V. Prokofev and B. V. Svistunov},
journal = {Physical Review Letters},
pages = {070601},
title = {Worm Algorithm for Continuous-Space Path Integral {M}onte {C}arlo Simulations},
volume = {96},
year = {2006},
url = {https://journals.aps.org/prl/abstract/10.1103/PhysRevLett.96.070601},
}

@article{boninsegni2006worm_b,
author = {M. Boninsegni and N. V. Prokofev and B. V. Svistunov},
journal = {Physical Review  E},
pages = {036701},
title = {Worm algorithm and diagrammatic {M}onte {C}arlo: A new approach to continuous-space path integral {M}onte {C}arlo simulations},
volume = {74},
year = {2006},
url = {https://journals.aps.org/pre/abstract/10.1103/PhysRevE.74.036701},
}

@article{boninsegni2017kinetic,
    author = {Boninsegni, Massimo},
    title = {Kinetic energy and momentum distribution of isotopic liquid helium mixtures},
    journal = {The Journal of Chemical Physics},
    volume = {148},
    number = {10},
    pages = {102308},
    year = {2017},
    month = {09},
    issn = {0021-9606},
    doi = {10.1063/1.5000101},
    url = {https://doi.org/10.1063/1.5000101}
}

@article{bonitz2020ab,
  title={\textit{Ab initio} simulation of warm dense matter},
  author={Bonitz, M and Dornheim, T and Moldabekov, Zh A and Zhang, S and Hamann, P and K{\"a}hlert, H and Filinov, A and Ramakrishna, K and Vorberger, J},
  journal={Physics of Plasmas},
  volume={27},
  number={4},
  pages={042710},
  year={2020},
  doi={10.1063/1.5143225},
  publisher={AIP Publishing LLC}
}

@article{bonitz2024principles,
  title={Toward first principles-based simulations of dense hydrogen},
  author={Michael Bonitz and Jan Vorberger and Mandy Bethkenhagen and Maximilian Böhme and David Ceperley and Alexey Filinov and Thomas Gawne and Frank Graziani and Gianluca Gregori and Paul Hamann and Stephanie Hansen and Markus Holzmann and S. X. Hu and Hanno Kählert and Valentin Karasiev and Uwe Kleinschmidt and Linda Kordts and Christopher Makait and Burkhard Militzer and Zhandos Moldabekov and Carlo Pierleoni and Martin Preising and Kushal Ramakrishna and Ronald Redmer and Sebastian Schwalbe and Pontus Svensson and Tobias Dornheim},
  journal={Physics of Plasmas},
  volume={31},
  number={11},
  pages={110501},
  year={2024},
  doi={10.1063/5.0219405},
  publisher={AIP Publishing}
}

@article{bowen1994static,
  title={Static dielectric response of the electron gas},
  author={Bowen, C and Sugiyama, G and Alder, BJ},
  journal={Physical Review B},
  volume={50},
  number={20},
  pages={14838},
  year={1994},
  publisher={APS}
}

@article{boronat1999quantum,
  title={Quantum {Monte} {Carlo} study of static properties of one $^3${He} atom in superfluid $^4${He}},
  author={Boronat, J and Casulleras, J},
  journal={Physical Review B},
  volume={59},
  number={13},
  pages={8844},
  year={1999},
  publisher={APS}
}

@article{brovman1974phonons,
  title={Phonons in nontransition metals},
  author={Brovman, Evgenii Grigor'evich and Kagan, Yurii M},
  journal={Soviet Physics Uspekhi},
  volume={17},
  number={2},
  pages={125},
  year={1974},
  publisher={IOP Publishing}
}

@inbook{brunton2022data,
  title={Data-driven science and engineering: {Machine} learning, dynamical systems, and control},
  author={Brunton, Steven L and Kutz, J Nathan},
  year={2022},
  chapter={4},
  publisher={Cambridge University Press}, 
}

@article{cenni1988evaluation,
  title={Evaluation of a class of diagrams useful in many-body calculations},
  author={Cenni, R and Saracco, P},
  journal={Nuclear Physics A},
  volume={487},
  number={2},
  pages={279--300},
  year={1988},
  publisher={Elsevier}
}

@incollection{ceperley1979quantum,
  title={Quantum many-body problems},
  author={Ceperley, David M and Kalos, M H},
  booktitle={{Monte} {Carlo} methods in statistical physics},
  pages={145--194},
  year={1979},
  publisher={Springer}
}

@Article{ceperley1991fermion,
author={Ceperley, D. M.},
title={Fermion nodes},
journal={Journal of Statistical Physics},
year={1991},
month={Jun},
day={01},
volume={63},
number={5},
pages={1237-1267},
issn={1572-9613},
doi={10.1007/BF01030009},
url={https://doi.org/10.1007/BF01030009}
}

@article{cenni1992fermionic,
  title={Fermionic loops in a many-body theory: {Evaluation} of an infinite class of diagrams},
  author={Cenni, R and Conte, F and Cornacchia, A and Saracco, P},
  journal={La Rivista del Nuovo Cimento},
  volume={15},
  number={12},
  pages={1--54},
  year={1992},
  publisher={Societ{\`a} Italiana di Fisica Bologna}
}

@article{ceperley1995path,
    author = {D. M. Ceperley},
    journal = {Reviews of Modern Physics},
    pages = {279},
    title = {Path integrals in the theory of condensed {Helium}},
    volume = {67},
    year = {1995},
    url = {https://journals.aps.org/rmp/abstract/10.1103/RevModPhys.67.279},
    doi={10.1103/RevModPhys.67.279}
}

@article{chabrier2000cooling,
  title={Cooling sequences and color-magnitude diagrams for cool white dwarfs with hydrogen atmospheres},
  author={Chabrier, Gilles and Brassard, Pierre and Fontaine, Gilles and Saumon, Didier},
  journal={The Astrophysical Journal},
  volume={543},
  number={1},
  pages={216},
  year={2000},
  doi={10.1086/317092},
  publisher={IOP Publishing}
}

@article{chiesa2006finte,
  title = {Finite-Size Error in Many-Body Simulations with Long-Range Interactions},
  author = {Chiesa, Simone and Ceperley, David M. and Martin, Richard M. and Holzmann, Markus},
  journal = {Physical Review Letters},
  volume = {97},
  issue = {7},
  pages = {076404},
  numpages = {4},
  year = {2006},
  month = {Aug},
  publisher = {American Physical Society},
  doi = {10.1103/PhysRevLett.97.076404},
  url = {https://link.aps.org/doi/10.1103/PhysRevLett.97.076404}
}

@article{chin2015high,
  title={High-order path-integral {Monte} {Carlo} methods for solving quantum dot problems},
  author={Chin, Siu A},
  journal={Physical Review E},
  volume={91},
  number={3},
  pages={031301},
  year={2015},
  publisher={APS}
}

@article{dalton2013linear,
  title={Linear and nonlinear density response functions for a simple atomic fluid},
  author={Dalton, Benjamin A and Glavatskiy, Kirill S and Daivis, Peter J and Todd, BD and Snook, Ian K},
  journal={The Journal of Chemical Physics},
  volume={139},
  number={4},
  year={2013},
  publisher={AIP Publishing}
}

@article{davis2020ion,
  title={Ion modes in dense ionized plasmas through nonadiabatic molecular dynamics},
  author={Davis, Ryan A and Angermeier, W A and Hermsmeier, R K T and White, Thomas G},
  journal={Physical Review Research},
  volume={2},
  number={4},
  pages={043139},
  year={2020},
  publisher={APS}
}

@article{dornheim2015permutation,
  title={Permutation blocking path integral {Monte} {Carlo} approach to the uniform electron gas at finite temperature},
  author={Dornheim, Tobias and Schoof, Tim and Groth, Simon and Filinov, Alexey and Bonitz, Michael},
  journal={The Journal of chemical physics},
  volume={143},
  number={20},
  year={2015},
  publisher={AIP Publishing}
}

@article{dornheim2016ab,
  title={\textit{Ab initio} quantum {Monte} {Carlo} simulation of the warm dense electron gas in the thermodynamic limit},
  author={Dornheim, Tobias and Groth, Simon and Sjostrom, Travis and Malone, Fionn D and Foulkes, WMC and Bonitz, Michael},
  journal={Physical Review Letters},
  volume={117},
  number={15},
  pages={156403},
  year={2016},
  publisher={APS}
}

@article{dornheim2018ab,
  title={\textit{Ab initio} path integral {Monte} {Carlo} results for the dynamic structure factor of correlated electrons: {From} the electron liquid to warm dense matter},
  author={Dornheim, Tobias and Groth, Simon and Vorberger, Jan and Bonitz, Michael},
  journal={Physical Review Letters},
  volume={121},
  number={25},
  pages={255001},
  year={2018},
  publisher={APS}
}

@article{dornheim2018uniform,
  title={The uniform electron gas at warm dense matter conditions},
  author={Dornheim, Tobias and Groth, Simon and Bonitz, Michael},
  journal={Physics Reports},
  volume={744},
  pages={1--86},
  year={2018},
  publisher={Elsevier}
}

@article{dornheim2019fermion,
author = {T. Dornheim},
journal = {Physical Review E},
pages = {023307},
title = {Fermion sign problem in path integral {M}onte {C}arlo simulations: {Q}uantum dots, ultracold atoms, and warm dense matter},
volume = {100},
year = {2019},
url = {https://journals.aps.org/pre/abstract/10.1103/PhysRevE.100.023307},
}

@article{dornheim2019static,
  title={The static local field correction of the warm dense electron gas: An \textit{ab} initio path integral {Monte} {Carlo} study and machine learning representation},
  author={Dornheim, Tobias and Vorberger, Jan and Groth, Simon and Hoffmann, Nico and Moldabekov, Zh A and Bonitz, Michael},
  journal={The Journal of Chemical Physics},
  volume={151},
  number={19},
  year={2019},
  publisher={AIP Publishing}
}

@article{dornheim2020effective,
  title={Effective static approximation: {A} fast and reliable tool for warm-dense matter theory},
  author={Dornheim, Tobias and Cangi, Attila and Ramakrishna, Kushal and B{\"o}hme, Maximilian and Tanaka, Shigenori and Vorberger, Jan},
  journal={Physical Review Letters},
  volume={125},
  number={23},
  pages={235001},
  year={2020},
  publisher={APS}
}

@article{dornheim2020nonlinear,
  title={Nonlinear electronic density response in warm dense matter},
  author={Dornheim, Tobias and Vorberger, Jan and Bonitz, Michael},
  journal={Physical Review Letters},
  volume={125},
  number={8},
  pages={085001},
  year={2020},
  publisher={APS}
}

@article{dornheim2021ab,
  title = {\textit{Ab initio} path integral {Monte} {Carlo} approach to the momentum distribution of the uniform electron gas at finite temperature without fixed nodes},
  author = {Dornheim, Tobias and B\"ohme, Maximilian and Militzer, Burkhard and Vorberger, Jan},
  journal = {Physical Review B},
  volume = {103},
  issue = {20},
  pages = {205142},
  numpages = {17},
  year = {2021},
  month = {May},
  publisher = {American Physical Society},
  doi = {10.1103/PhysRevB.103.205142},
  url = {https://link.aps.org/doi/10.1103/PhysRevB.103.205142}
}

@article{dornheim2021density,
  title={Density response of the warm dense electron gas beyond linear response theory: {Excitation} of harmonics},
  author={Dornheim, Tobias and B{\"o}hme, Maximilian and Moldabekov, Zhandos A and Vorberger, Jan and Bonitz, Michael},
  journal={Physical Review Research},
  volume={3},
  number={3},
  pages={033231},
  year={2021},
  publisher={APS}
}

@article{dornheim2021nonlinear,
  title={Nonlinear density response from imaginary-time correlation functions: {Ab} initio path integral {Monte} {Carlo} simulations of the warm dense electron gas},
  author={Dornheim, Tobias and Moldabekov, Zhandos A and Vorberger, Jan},
  journal={The Journal of Chemical Physics},
  volume={155},
  number={5},
  year={2021},
  publisher={AIP Publishing}
}

@article{dornheim2021overcoming,
  title={Overcoming finite-size effects in electronic structure simulations at extreme conditions},
  author={Dornheim, Tobias and Vorberger, Jan},
  journal={The Journal of Chemical Physics},
  volume={154},
  number={14},
  year={2021},
  publisher={AIP Publishing}
}

@article{dornheim2022electronic,
  title={Electronic pair alignment and roton feature in the warm dense electron gas},
  author={Dornheim, Tobias and Moldabekov, Zhandos and Vorberger, Jan and K{\"a}hlert, Hanno and Bonitz, Michael},
  journal={Communications Physics},
  volume={5},
  number={1},
  pages={304},
  year={2022},
  publisher={Nature Publishing Group UK London}
}

@article{dornheim2022nonlinear,
  title={Nonlinear interaction of external perturbations in warm dense matter},
  author={Dornheim, Tobias and Vorberger, Jan and Moldabekov, Zhandos A and Bonitz, Michael},
  journal={Contributions to Plasma Physics},
  volume={62},
  number={10},
  pages={e202100247},
  year={2022},
  publisher={Wiley Online Library}
}

@Article{dornheim2022path,
author={Dornheim, Tobias
and Moldabekov, Zhandos A.
and Vorberger, Jan
and Militzer, Burkhard},
title="{Path integral {Monte Carlo} approach to the structural properties and collective excitations of liquid $^3$He without fixed nodes}",
journal={Scientific Reports},
year={2022},
month={Jan},
day={13},
volume={12},
number={1},
pages={708},
issn={2045-2322},
doi={10.1038/s41598-021-04355-9},
url={https://doi.org/10.1038/s41598-021-04355-9}
}

@article{dornheim2022spin,
  title={Spin-resolved density response of the warm dense electron gas},
  author={Dornheim, Tobias and Vorberger, Jan and Moldabekov, Zhandos A and Tolias, Panagiotis},
  journal={Physical Review Research},
  volume={4},
  number={3},
  pages={033018},
  year={2022},
  publisher={APS}
}

@article{dornheim2023electronic,
    author = {Dornheim, Tobias and Moldabekov, Zhandos A. and Ramakrishna, Kushal and Tolias, Panagiotis and Baczewski, Andrew D. and Kraus, Dominik and Preston, Thomas R. and Chapman, David A. and Böhme, Maximilian P. and Döppner, Tilo and Graziani, Frank and Bonitz, Michael and Cangi, Attila and Vorberger, Jan},
    title = "{Electronic density response of warm dense matter}",
    journal = {Physics of Plasmas},
    volume = {30},
    number = {3},
    year = {2023},
    pages={032705},
    month = {03},
    issn = {1070-664X},
    doi = {10.1063/5.0138955},
    url = {https://doi.org/10.1063/5.0138955}
}

@article{dornheim2023energy,
  title={Energy response and spatial alignment of the perturbed electron gas},
  author={Dornheim, Tobias and Tolias, Panagiotis and Moldabekov, Zhandos A and Vorberger, Jan},
  journal={The Journal of Chemical Physics},
  volume={158},
  number={16},
  year={2023},
  publisher={AIP Publishing}
}

@article{dornheim2024ab,
  title={\textit{Ab initio} density response and local field factor of warm dense hydrogen},
  author={Dornheim, Tobias and Schwalbe, Sebastian and Tolias, Panagiotis and B{\"o}hme, Maximilian P and Moldabekov, Zhandos A and Vorberger, Jan},
  journal={Matter and Radiation at Extremes},
  volume={9},
  number={5},
  year={2024},
  publisher={AIP Publishing}
}

@article{dornheim2025direct,
  title={Direct free energy calculation from \textit{ab initio} path integral {Monte} {Carlo} simulations of warm dense matter},
  author={Dornheim, Tobias and Moldabekov, Zhandos A and Schwalbe, Sebastian and Vorberger, Jan},
  journal={Physical Review B},
  volume={111},
  number={4},
  pages={L041114},
  year={2025},
  publisher={APS},
  doi={10.1103/PhysRevB.111.L041114}
}

@article{dornheim2025eta,
  title={$\eta$-ensemble path integral {Monte} {Carlo} approach to the free energy of the warm dense electron gas and the uniform electron liquid},
  author={Dornheim, Tobias and Tolias, Panagiotis and Moldabekov, Zhandos A and Vorberger, Jan},
  journal={Physical Review Research},
  volume={7},
  number={2},
  pages={023250},
  year={2025},
  publisher={APS},
  doi={10.1103/4n7x-78fs}
}

@Article{dornheim2025fermionic,
author={Dornheim, Tobias
and Moldabekov, Zhandos
and Schwalbe, Sebastian
and Tolias, Panagiotis
and Vorberger, Jan},
title={Fermionic Free Energies from \textit{Ab Initio} Path Integral {Monte Carlo} Simulations of Fictitious Identical Particles},
journal={Journal of Chemical Theory and Computation},
year={2025},
month={Aug},
day={12},
publisher={American Chemical Society},
volume={21},
number={15},
pages={7290-7303},
issn={1549-9618},
doi={10.1021/acs.jctc.5c00301},
url={https://doi.org/10.1021/acs.jctc.5c00301}
}

@article{dornheim2025reweighting,
  title={Reweighting estimator for \textit{ab initio} path integral {Monte} {Carlo} simulations of fictitious identical particles},
  author={Dornheim, Tobias and Svensson, Pontus and Hamann, Paul and Schwalbe, Sebastian and Moldabekov, Zhandos A and Tolias, Panagiotis and Vorberger, Jan},
  journal={The Journal of Chemical Physics},
  volume={163},
  number={15},
  year={2025},
  publisher={AIP Publishing}
}

@article{dornheim2025unraveling,
  title={Unraveling electronic correlations in warm dense quantum plasmas},
  author={Dornheim, Tobias and D{\"o}ppner, Tilo and Tolias, Panagiotis and B{\"o}hme, Maximilian P and Fletcher, Luke B and Gawne, Thomas and Graziani, Frank R and Kraus, Dominik and MacDonald, Michael J and Moldabekov, Zhandos A and others},
  journal={Nature Communications},
  volume={16},
  number={1},
  pages={5103},
  year={2025},
  publisher={Nature Publishing Group UK London}
}

@article{farid1993extremal,
  title={Extremal properties of the {Harris-Foulkes} functional and an improved screening calculation for the electron gas},
  author={Farid, Behnam and Heine, Volker and Engel, G E and Robertson, I J},
  journal={Physical Review B},
  volume={48},
  number={16},
  pages={11602},
  year={1993},
  publisher={APS}
}

@article{ferrenberg1988new,
  title={New {Monte} {Carlo} technique for studying phase transitions},
  author={Ferrenberg, Alan M and Swendsen, Robert H},
  journal={Physical Review Letters},
  volume={61},
  number={23},
  pages={2635},
  year={1988},
  publisher={APS}
}

@article{ferre2016dynamic,
  title = {Dynamic structure factor of liquid $^{4}\mathrm{He}$ across the normal-superfluid transition},
  author = {Ferr\'e, G. and Boronat, J.},
  journal = {Physical Review B},
  volume = {93},
  issue = {10},
  pages = {104510},
  numpages = {9},
  year = {2016},
  month = {Mar},
  publisher = {American Physical Society},
  doi = {10.1103/PhysRevB.93.104510},
  url = {https://link.aps.org/doi/10.1103/PhysRevB.93.104510}
}

@article{fletcher2022electron,
  title={Electron-ion temperature relaxation in warm dense hydrogen observed with picosecond resolved x-ray scattering},
  author={Fletcher, Luke B and Vorberger, Jan and Schumaker, Will and Ruyer, Charles and Goede, Sebastian and Galtier, Eric and Zastrau, Ulf and Alves, Eduardo P and Baalrud, Scott D and Baggott, Rory A and others},
  journal={Frontiers in Physics},
  volume={10},
  pages={838524},
  year={2022},
  publisher={Frontiers Media SA},
  doi={10.3389/fphy.2022.838524}
}

@article{filinov2010berezinskii,
  title = {{Berezinskii-Kosterlitz-Thouless} Transition in Two-Dimensional Dipole Systems},
  author = {Filinov, A. and Prokof'ev, N. V. and Bonitz, M.},
  journal = {Physical Review Letters},
  volume = {105},
  issue = {7},
  pages = {070401},
  numpages = {4},
  year = {2010},
  month = {Aug},
  publisher = {American Physical Society},
  doi = {10.1103/PhysRevLett.105.070401},
  url = {https://link.aps.org/doi/10.1103/PhysRevLett.105.070401}
}

@article{filinov2012collective,
  title = {Collective and single-particle excitations in two-dimensional dipolar Bose gases},
  author = {Filinov, A. and Bonitz, M.},
  journal = {Physical Review A},
  volume = {86},
  issue = {4},
  pages = {043628},
  numpages = {26},
  year = {2012},
  month = {Oct},
  publisher = {American Physical Society},
  doi = {10.1103/PhysRevA.86.043628},
  url = {https://link.aps.org/doi/10.1103/PhysRevA.86.043628}
}

@article{filinov2013color,
  title = {Color path-integral {Monte-Carlo} simulations of quark-gluon plasma: {Thermodynamic} and transport properties},
  author = {Filinov, V. S. and Ivanov, Yu. B. and Fortov, V. E. and Bonitz, M. and Levashov, P. R.},
  journal = {Physical Review C},
  volume = {87},
  issue = {3},
  pages = {035207},
  numpages = {20},
  year = {2013},
  month = {Mar},
  publisher = {American Physical Society},
  doi = {10.1103/PhysRevC.87.035207},
  url = {https://link.aps.org/doi/10.1103/PhysRevC.87.035207}
}

@article{fortney2010interior,
  title={The Interior Structure, Composition, and Evolution of Giant Planets},
  author={Fortney, Jonathan J and Nettelmann, Nadine},
  journal={Space Science Reviews},
  volume={152},
  pages={423--447},
  year={2010},
  doi={10.1007/s11214-009-9582-x},
  publisher={Springer}
}

@article{fortov2009extreme,
  title={Extreme states of matter on Earth and in space},
  author={Fortov, Vladimir E},
  journal={Physics-Uspekhi},
  volume={52},
  number={6},
  pages={615},
  year={2009},
  doi={10.3367/UFNe.0179.200906h.0653},
  publisher={IOP Publishing}
}

@article{french2012ab,
  title={\textit{Ab initio} simulations for material properties along the {Jupiter} adiabat},
  author={French, Martin and Becker, Andreas and Lorenzen, Winfried and Nettelmann, Nadine and Bethkenhagen, Mandy and Wicht, Johannes and Redmer, Ronald},
  journal={The Astrophysical Journal Supplement Series},
  volume={202},
  number={1},
  pages={5},
  year={2012},
  doi={10.1088/0067-0049/202/1/5},
  publisher={IOP Publishing}
}

@inbook{frenkel2002understanding,
  title={Understanding Molecular Simulation: {From} Algorithms to Applications},
  chapter={Free Energy Calculations},
  pages={167--200},
  author={Frenkel, Daan and Smit, Berend},
  edition={2},
  number={2.2},
  year={2002},
  publisher={Academic Press},
  address={San Diego, USA}
}

@book{giuliani2008quantum,
  title={Quantum {T}heory of the {E}lectron {L}iquid},
  author={Giuliani, Gabriele and Vignale, Giovanni},
  year={2008},
  publisher={Cambridge University Press},
  address={Cambridge, UK}  
}

@article{glenzer2009x,
  title={X-ray {Thomson} scattering in high energy density plasmas},
  author={Glenzer, Siegfried H. and Redmer, Ronald},
  journal={Reviews of Modern Physics},
  volume={81},
  issue={4},
  pages={1625},
  year={2009},
  month={12},
  publisher={American Physical Society},
  doi={10.1103/RevModPhys.81.1625},
  url={https://link.aps.org/doi/10.1103/RevModPhys.81.1625}
}

@article{gudmundsson1983structure,
  title={Structure of neutron star envelopes},
  author={Gudmundsson, Einar H and Pethick, C J and Epstein, Richard I},
  journal={Astrophysical Journal},
  volume={272},
  pages={286--300},
  doi={10.1086/161292},
  year={1983}
}

@article{guillot1999interiors,
  title={Interiors of giant planets inside and outside the solar system},
  author={Guillot, Tristan},
  journal={Science},
  volume={286},
  number={5437},
  pages={72--77},
  year={1999},
  doi={10.1126/science.286.5437.72},
  publisher={American Association for the Advancement of Science}
}

@article{gravel2007nonlinear,
  title={Nonlinear response theories and effective pair potentials},
  author={Gravel, Simon and Ashcroft, N W},
  journal={Physical Review B},
  volume={76},
  number={14},
  pages={144103},
  year={2007},
  publisher={APS}
}

@book{graziani2014frontiers,
address = {International Publishing},
editor = {F. Graziani and M. P. Desjarlais and R. Redmer and S. B. Trickey},
publisher = {Springer},
title = {Frontiers and Challenges in Warm Dense Matter},
year = {2014},
}

@article{groth2019ab,
  title={\textit{Ab initio} path integral {Monte} {Carlo} approach to the static and dynamic density response of the uniform electron gas},
  author={Groth, Simon and Dornheim, Tobias and Vorberger, Jan},
  journal={Physical Review B},
  volume={99},
  number={23},
  pages={235122},
  year={2019},
  publisher={APS}
}

@article{hamann2023prediction,
  title = {Prediction of a roton-type feature in warm dense hydrogen},
  author = {Hamann, Paul and Kordts, Linda and Filinov, Alexey and Bonitz, Michael and Dornheim, Tobias and Vorberger, Jan},
  journal = {Physical Review Research},
  volume = {5},
  issue = {3},
  pages = {033039},
  numpages = {11},
  year = {2023},
  month = {Jul},
  publisher = {American Physical Society},
  doi = {10.1103/PhysRevResearch.5.033039},
  url = {https://link.aps.org/doi/10.1103/PhysRevResearch.5.033039}
}

@article{hamann2026reweighting,
    author = {Hamann, Paul and Vorberger, Jan and Dornheim, Tobias},
    title = {Reweighting Scheme for the Calculation of Grand-Canonical Expectation Values in Quantum {Monte} {Carlo} Simulations With a {Fermion} Sign Problem},
    journal = {Contributions to Plasma Physics},
    pages = {e70091},
    year={2026},
    doi = {https://doi.org/10.1002/ctpp.70091},
    url = {https://onlinelibrary.wiley.com/doi/abs/10.1002/ctpp.70091}
}

@book{hansen1993theory,
  title={Theory of {S}imple {L}iquids},
  author={Hansen, Jean-Pierre and McDonald, Ian Ranald},
  year={1986},
  edition={2nd ed.},
  publisher={Academic Press},
  address={London}
}

@book{haensel2007neutron,
  title={Neutron {S}tars 1: {Equation} of {S}tate and {S}tructure},
  author={Haensel, Pawe{\l} and Potekhin, Aleksander Yu and Yakovlev, Dmitry G},
  year={2007},
  publisher={Springer},
  address={New York, NY}
}

@article{hatano1994data,
  title={Data analysis for quantum {Monte} {Carlo} simulations with the negative-sign problem},
  author={Hatano, Naomichi},
  journal={Journal of the Physical Society of Japan},
  volume={63},
  number={5},
  pages={1691--1697},
  year={1994},
  publisher={The Physical Society of Japan},
  doi={10.1143/JPSJ.63.1691}
}

@article{helled2020understanding,
  title={Understanding dense hydrogen at planetary conditions},
  author={Helled, Ravit and Mazzola, Guglielmo and Redmer, Ronald},
  journal={Nature Reviews Physics},
  volume={2},
  number={10},
  pages={562--574},
  year={2020},
  doi={10.1038/s42254-020-0223-3},
  publisher={Nature Publishing Group UK London}
}

@article{hentschel2025statistical,
  title={Statistical inference of collision frequencies from x-ray {Thomso}n scattering spectra},
  author={Hentschel, Thomas W and Kononov, Alina and Baczewski, Andrew D and Hansen, Stephanie B},
  journal={Physics of Plasmas},
  volume={32},
  number={1},
  year={2025},
  publisher={AIP Publishing}
}

@article{hou2022exchange,
  title={Exchange-correlation effect in the charge response of a warm dense electron gas},
  author={Hou, Peng-Cheng and Wang, Bao-Zong and Haule, Kristjan and Deng, Youjin and Chen, Kun},
  journal={Physical Review B},
  volume={106},
  number={8},
  pages={L081126},
  year={2022},
  publisher={APS}
}

@article{hu1988z,
  title={{$Z^3$} correction to the stopping power of ions in an electron gas},
  author={Hu, C D and Zaremba, E},
  journal={Physical review B},
  volume={37},
  number={16},
  pages={9268},
  year={1988},
  publisher={APS}
}

@article{hurricane2023physics,
  title={Physics principles of inertial confinement fusion and {US} program overview},
  author={Hurricane, O A and Patel, P K and Betti, R and Froula, D H and Regan, S P and Slutz, S A and Gomez, M R and Sweeney, M A},
  journal={Reviews of Modern Physics},
  volume={95},
  number={2},
  pages={025005},
  year={2023},
  doi={10.1103/RevModPhys.95.025005},
  publisher={APS}
}

@software{ISHTAR,
  author       = {Dornheim, Tobias and
                  Böhme, Maximilian and
                  Schwalbe, Sebastian},
  title        = {{ISHTAR - Imaginary-time Stochastic High- 
                   performance Tool for \textit{Ab initio} Research}},
  month        = jan,
  year         = 2024,
  publisher    = {Zenodo},
  doi          = {10.5281/zenodo.10497098},
  url          = {https://doi.org/10.5281/zenodo.10497098}
}

@article{kahlert2026structural,
  title={Structural and dynamic properties of dense hydrogen plasmas from semi-classical molecular dynamics simulations},
  author={K{\"a}hlert, Hanno},
  journal={Physics of Plasmas},
  volume={33},
  number={2},
  year={2026},
  publisher={AIP Publishing}
}

@article{karmakar2026combining,
  title={Combining harmonic sampling with the worm algorithm to improve the efficiency of path integral {Monte} {Carlo}},
  author={Karmakar, Sourav and Paul, Sutirtha and Del Maestro, Adrian and Hirshberg, Barak},
  journal={Physical Review E},
  volume={113},
  number={2},
  pages={025306},
  year={2026},
  publisher={APS}
}

@book{kippenhahn2012stellar,
  title={Stellar {S}tructure and {E}volution},
  author={Kippenhahn, Rudolf and Weigert, Alfred and Weiss, Achim},
  edition={2 ed.},
  year={2012},
  publisher={Springer},
  address={Heidelberg}
}

@misc{koskelo2023shortrange,
      title={Short-range excitonic phenomena in low-density metals}, 
      author={Jaakko Koskelo and Lucia Reining and Matteo Gatti},
      year={2023},
      eprint={2301.00474},
      archivePrefix={arXiv},
      primaryClass={cond-mat.str-el}
}

@article{kubo1966fluctuation,
  title={The fluctuation-dissipation theorem},
  author={Kubo, Rep},
  journal={Reports on progress in physics},
  volume={29},
  number={1},
  pages={255--284},
  year={1966}
}

@book{landau2021guide, 
  address={Cambridge}, 
  edition={5}, 
  title={A Guide to {Monte} {Carlo} Simulations in Statistical Physics}, 
  publisher={Cambridge University Press}, 
  author={Landau, David and Binder, Kurt}, 
  year={2021}
}

@article{larder2019fast,
  title={Fast nonadiabatic dynamics of many-body quantum systems},
  author={Larder, Brett and Gericke, Dirk O and Richardson, Scott and Mabey, Paul and White, TG and Gregori, Gianluca},
  journal={Science advances},
  volume={5},
  number={11},
  pages={eaaw1634},
  year={2019},
  publisher={American Association for the Advancement of Science}
}

@article{lloyd1968structural,
  title={A structural expansion of the cohesive energy of simple metals in an effective {Hamiltonian} approximation},
  author={Lloyd, P and Sholl, C A},
  journal={Journal of Physics C},
  volume={1},
  number={6},
  pages={1620--1632},
  year={1968}
}

@article{loh1990sign,
  title={Sign problem in the numerical simulation of many-electron systems},
  author={Loh Jr, E Y and Gubernatis, J E and Scalettar, R T and White, S R and Scalapino, D J and Sugar, R L},
  journal={Physical Review B},
  volume={41},
  number={13},
  pages={9301},
  year={1990},
  publisher={APS}
}

@article{louis1998extending,
  title={Extending linear response: {Inferences} from electron-ion structure factors},
  author={Louis, A A and Ashcroft, N W},
  journal={Physical Review Letters},
  volume={81},
  number={20},
  pages={4456},
  year={1998},
  publisher={APS}
}

@article{mcdonald1967calculation,
  title={Calculation of thermodynamic properties of liquid argon from {Lennard-Jones} parameters by a {Monte} {Carlo} method},
  author={McDonald, Ian R and Singer, Konrad},
  journal={Discussions of the Faraday Society},
  volume={43},
  pages={40--49},
  year={1967},
  publisher={Royal Society of Chemistry}
}

@article{miao2020chemistry,
  title={Chemistry under high pressure},
  author={Miao, Maosheng and Sun, Yuanhui and Zurek, Eva and Lin, Haiqing},
  journal={Nature Reviews Chemistry},
  volume={4},
  number={10},
  pages={508--527},
  year={2020},
  publisher={Nature Publishing Group UK London},
  doi={10.1038/s41570-020-0213-0}
}

@article{mikhailov2012second,
  title={Second-order response of a uniform three-dimensional electron gas to a longitudinal electric field},
  author={Mikhailov, Sergey A},
  journal={Annalen der Physik},
  volume={524},
  number={3-4},
  pages={182--187},
  year={2012},
  publisher={Wiley Online Library}
}

@article{mikhailov2014nonlinear,
  title={Nonlinear electromagnetic response of a uniform electron gas},
  author={Mikhailov, Sergey A},
  journal={Physical Review Letters},
  volume={113},
  number={2},
  pages={027405},
  year={2014},
  publisher={APS}
}

@article{milchev1977quadratic,
  title={The quadratic response of a fermi gas. {Building-up} of the covalent bonding in zineblende semiconductors. {A} nondiagonal density matrix treatment},
  author={Milchev, A and Pickenhain, R},
  journal={Physica status solidi (b)},
  volume={79},
  number={2},
  pages={549--558},
  year={1977},
  publisher={Wiley Online Library}
}

@article{moldabekov2018theoretical,
  title={Theoretical foundations of quantum hydrodynamics for plasmas},
  author={Moldabekov, Zh A and Bonitz, M and Ramazanov, T S},
  journal={Physics of Plasmas},
  volume={25},
  number={3},
  year={2018},
  publisher={AIP Publishing}
}

@article{moldabekov2021relevance,
  title={The relevance of electronic perturbations in the warm dense electron gas},
  author={Moldabekov, Zhandos and Dornheim, Tobias and B{\"o}hme, Maximilian and Vorberger, Jan and Cangi, Attila},
  journal={The Journal of Chemical Physics},
  volume={155},
  number={12},
  year={2021},
  publisher={AIP Publishing}
}

@article{moldabekov2022benchmarking,
  title={Benchmarking exchange-correlation functionals in the spin-polarized inhomogeneous electron gas under warm dense conditions},
  author={Moldabekov, Zhandos and Dornheim, Tobias and Vorberger, Jan and Cangi, Attila},
  journal={Physical Review B},
  volume={105},
  number={3},
  pages={035134},
  year={2022},
  publisher={APS}
}

@article{moldabekov2022density,
  title={Density functional theory perspective on the nonlinear response of correlated electrons across temperature regimes},
  author={Moldabekov, Zhandos and Vorberger, Jan and Dornheim, Tobias},
  journal={Journal of Chemical Theory and Computation},
  volume={18},
  number={5},
  pages={2900--2912},
  year={2022},
  publisher={ACS Publications}
}

@article{moldabekov2023assessing,
  title={Assessing the accuracy of hybrid exchange-correlation functionals for the density response of warm dense electrons},
  author={Moldabekov, Zhandos A and Lokamani, Mani and Vorberger, Jan and Cangi, Attila and Dornheim, Tobias},
  journal={The Journal of Chemical Physics},
  volume={158},
  number={9},
  year={2023},
  publisher={AIP Publishing}
}

@article{moldabekov2023linear,
  title={Linear-response time-dependent density functional theory approach to warm dense matter with adiabatic exchange-correlation kernels},
  author={Moldabekov, Zhandos A and Pavanello, Michele and B{\"o}hme, Maximilian P and Vorberger, Jan and Dornheim, Tobias},
  journal={Physical Review Research},
  volume={5},
  number={2},
  pages={023089},
  year={2023},
  publisher={APS}
}

@article{moldabekov2023imposing,
  title={Imposing correct jellium response is key to predict the density response by orbital-free {DFT}},
  author={Moldabekov, Zhandos A and Shao, Xuecheng and Pavanello, Michele and Vorberger, Jan and Graziani, Frank and Dornheim, Tobias},
  journal={Physical Review B},
  volume={108},
  number={23},
  pages={235168},
  year={2023},
  publisher={APS}
}

@article{moldabekov2025applying,
  title={Applying the {Liouville--Lanczos} method of time-dependent density-functional theory to warm dense matter},
  author={Moldabekov, Zhandos A and Schwalbe, Sebastian and Gawne, Thomas and Preston, Thomas R and Vorberger, Jan and Dornheim, Tobias},
  journal={Matter and Radiation at Extremes},
  volume={10},
  number={4},
  year={2025},
  publisher={AIP Publishing}
}

@article{moldabekov2025density,
  title={From density response to energy functionals and back: {An} \textit{ab initio} perspective on matter under extreme conditions},
  author={Moldabekov, Zhandos and Vorberger, Jan and Dornheim, Tobias},
  journal={Progress in Particle and Nuclear Physics},
  volume={140},
  pages={104144},
  year={2025},
  publisher={Elsevier}
}

@article{moldabekov2025generalized,
  title={Generalized density functional theory framework for the nonlinear density response of quantum many-body systems},
  author={Moldabekov, Zhandos A and Ma, Cheng and Shao, Xuecheng and Schwalbe, Sebastian and Svensson, Pontus and Tolias, Panagiotis and Vorberger, Jan and Dornheim, Tobias},
  journal={Physical Review B},
  volume={113},
  number={12},
  pages={125115},
  year={2026},
  publisher={APS}
}

@article{moldabekov2025enhancing,
  title={Enhancing the Efficiency of Time-Dependent Density Functional Theory Calculations of Dynamic Response Properties},
  author={Moldabekov, Zhandos A and Schwalbe, Sebastian and Acosta, Uwe Hernandez and Gawne, Thomas and Vorberger, Jan and Pavanello, Michele and Dornheim, Tobias},
  journal={arXiv preprint arXiv:2510.01875},
  year={2025}
}

@article{moroni1992static,
  title={Static response from quantum {Monte} {Carlo} calculations},
  author={Moroni, Saverio and Ceperley, David M and Senatore, Gaetano},
  journal={Physical Review Letters},
  volume={69},
  number={13},
  pages={1837},
  year={1992},
  publisher={APS}
}

@article{moroni1995static,
  title={Static response and local field factor of the electron gas},
  author={Moroni, Saverio and Ceperley, David M and Senatore, Gaetano},
  journal={Physical Review Letters},
  volume={75},
  number={4},
  pages={689},
  year={1995},
  publisher={APS}
}

@article{nagao2003enhanced,
  title={Enhanced {Friedel} structure and proton pairing in dense solid hydrogen},
  author={Nagao, Kazutaka and Bonev, S A and Bergara, A and Ashcroft, N W},
  journal={Physical Review Letters},
  volume={90},
  number={3},
  pages={035501},
  year={2003},
  publisher={APS}
}

@article{nuckolls1972laser,
  title={Laser compression of matter to super-high densities: {Thermonuclear} ({CTR}) applications},
  author={Nuckolls, John and Wood, Lowell and Thiessen, Albert and Zimmerman, George},
  journal={Nature},
  volume={239},
  number={5368},
  pages={139--142},
  year={1972},
  doi={10.1038/239139a0},
  publisher={Nature Publishing Group}
}

@Article{ott2018recent,
author={Ott, Torben
and Thomsen, Hauke
and Abraham, Jan Willem
and Dornheim, Tobias
and Bonitz, Michael},
title={Recent progress in the theory and simulation of strongly correlated plasmas: {Phase} transitions, transport, quantum, and magnetic field effects},
journal={The European Physical Journal D},
year={2018},
month={May},
day={22},
volume={72},
number={5},
pages={84},
issn={1434-6079},
doi={10.1140/epjd/e2018-80385-7},
url={https://doi.org/10.1140/epjd/e2018-80385-7}
}

@article{paasch1977quadratic,
  title={Quadratic electronic polarizability of the interacting electron gas},
  author={Paasch, G and Rennert, P},
  journal={physica status solidi (b)},
  volume={83},
  number={2},
  pages={501--508},
  year={1977},
  publisher={Wiley Online Library}
}

@article{paasch1980quadratic,
  title={Quadratic response of the electron gas. {Inclusion} of local field corrections},
  author={Paasch, G and Heinrich, A},
  journal={physica status solidi (b)},
  volume={102},
  number={1},
  pages={323--330},
  year={1980},
  publisher={Wiley Online Library}
}

@article{perdew2008restoring,
  title = {Restoring the Density-Gradient Expansion for Exchange in Solids and Surfaces},
  author = {Perdew, John P. and Ruzsinszky, Adrienn and Csonka, G\'abor I. and Vydrov, Oleg A. and Scuseria, Gustavo E. and Constantin, Lucian A. and Zhou, Xiaolan and Burke, Kieron},
  journal = {Physical Review Letters},
  volume = {100},
  issue = {13},
  pages = {136406},
  numpages = {4},
  year = {2008},
  month = {Apr},
  publisher = {American Physical Society},
  doi = {10.1103/PhysRevLett.100.136406},
  url = {https://link.aps.org/doi/10.1103/PhysRevLett.100.136406}
}

@book{pines2018theory,
  title={Theory of {Q}uantum {L}iquids: {Normal} {Fermi} liquids},
  author={Pines, David and Nozières, Philippe},
  year={2018},
  publisher={CRC Press},
  address={Boca Raton, FL}
}

@article{pribram2016thermal,
author = {A. Pribram-Jones and P. E. Grabowski and K. Burke},
journal = {Physical Review Letters},
pages = {233001},
title = {Thermal Density Functional Theory: {Time}-Dependent Linear Response and Approximate Functionals from the Fluctuation-Dissipation Theorem},
volume = {116},
year = {2016},
url = {https://journals.aps.org/prl/abstract/10.1103/PhysRevLett.116.233001},
}

@article{pitarke1995quadratic,
  title={Quadratic response theory of the energy loss of charged particles in an electron gas},
  author={Pitarke, J M and Ritchie, R H and Echenique, P M},
  journal={Physical Review B},
  volume={52},
  number={19},
  pages={13883},
  year={1995},
  publisher={APS}
}

@article{poole2024multimessenger,
  title = {Multimessenger measurements of the static structure of shock-compressed liquid silicon at 100 {GPa}},
  author = {Poole, H. and Ginnane, M. K. and Millot, M. and Bellenbaum, H. M. and Collins, G. W. and Hu, S. X. and Polsin, D. and Saha, R. and Topp-Mugglestone, J. and White, T. G. and Chapman, D. A. and Rygg, J. R. and Regan, S. P. and Gregori, G.},
  journal = {Physical Review Research},
  volume = {6},
  issue = {2},
  pages = {023144},
  numpages = {15},
  year = {2024},
  month = {May},
  publisher = {American Physical Society},
  doi = {10.1103/PhysRevResearch.6.023144},
  url = {https://link.aps.org/doi/10.1103/PhysRevResearch.6.023144}
}

@article{porter2010pair,
  title={Pair potentials for simple metallic systems: {Beyond} linear response},
  author={Porter, J A and Ashcroft, N W and Chester, G V},
  journal={Physical Review B},
  volume={81},
  number={22},
  pages={224113},
  year={2010},
  publisher={APS}
}

@article{preising2023material,
  title={Material properties of {Saturn’s} interior from \textit{ab initio} simulations},
  author={Preising, Martin and French, Martin and Mankovich, Christopher and Soubiran, Fran{\c{c}}ois and Redmer, Ronald},
  journal={The Astrophysical Journal Supplement Series},
  volume={269},
  number={2},
  pages={47},
  year={2023},
  doi={10.3847/1538-4365/ad0293},
  publisher={IOP Publishing}
}

@article{ritchie1959interaction,
  title={Interaction of charged particles with a degenerate {Fermi-Dirac} electron gas},
  author={Ritchie, R H},
  journal={Physical Review},
  volume={114},
  number={3},
  pages={644},
  year={1959},
  publisher={APS}
}

@incollection{rommel1998quadratic,
  title={The Quadratic Susceptibility in One, Two, and Three Dimensions},
  author={Rommel, J Martin and Kalman, Gabor J and Genga, Riewa},
  booktitle={Strongly Coupled Coulomb Systems},
  pages={669--672},
  year={1998},
  publisher={Springer}
}

@article{sakkos2009high,
  title={High order {Chin} actions in path integral {Monte} {Carlo}},
  author={Sakkos, K and Casulleras, J and Boronat, J},
  journal={The Journal of Chemical Physics},
  volume={130},
  number={20},
  year={2009},
  publisher={AIP Publishing}
}

@article{sakko2010time,
  title={Time-dependent density functional approach for the calculation of inelastic x-ray scattering spectra of molecules},
  author={Sakko, Arto and Rubio, Angel and Hakala, Mikko and H{\"a}m{\"a}l{\"a}inen, Keijo},
  journal={The Journal of Chemical Physics},
  volume={133},
  number={17},
  pages={174111},
  year={2010},
  doi={10.1063/1.3503594},
  publisher={AIP Publishing}
}

@article{schorner2022extending,
  title={Extending \textit{ab initio} simulations for the ion-ion structure factor of warm dense aluminum to the hydrodynamic limit using neural network potentials},
  author={Sch{\"o}rner, Maximilian and R{\"u}ter, Hannes R and French, Martin and Redmer, Ronald},
  journal={Physical Review B},
  volume={105},
  number={17},
  pages={174310},
  year={2022},
  publisher={APS}
}

@article{schorner2023x,
  title={X-ray {Thomson} scattering spectra from density functional theory molecular dynamics simulations based on a modified {Chihara} formula},
  author={Sch{\"o}rner, Maximilian and Bethkenhagen, Mandy and D{\"o}ppner, Tilo and Kraus, Dominik and Fletcher, Luke B and Glenzer, Siegfried H and Redmer, Ronald},
  journal={Physical Review E},
  volume={107},
  number={6},
  pages={065207},
  year={2023},
  publisher={APS}
}

@misc{schwalbe2025static,
      title={Static linear density response from X-ray {Thomson} scattering measurements: {A} case study of warm dense beryllium}, 
      author={Sebastian Schwalbe and Hannah Bellenbaum and Tilo Döppner and Maximilian Böhme and Thomas Gawne and Dominik Kraus and Michael J. MacDonald and Zhandos Moldabekov and Panagiotis Tolias and Jan Vorberger and Tobias Dornheim},
      year={2025},
      eprint={2504.13611},
      archivePrefix={arXiv},
      primaryClass={physics.plasm-ph},
      url={https://arxiv.org/abs/2504.13611}, 
}

@book{sitenko1982fluctuations,
  title={Fluctuations and Non-Linear Wave Interactions in Plasmas: {International} Series in Natural Philosophy},
  author={Sitenko, Alekse{\u\i} Grigorʹevich},
  volume={107},
  year={1982},
  publisher={Pergamon Press},
  address={Oxford}
}

@article{sturm1993dynamic,
  title={Dynamic structure factor: {An} introduction},
  author={Sturm, K},
  journal={Zeitschrift f{\"u}r Naturforschung A},
  volume={48},
  number={1-2},
  pages={233--242},
  year={1993},
  doi={10.1515/zna-1993-1-244},
  publisher={Verlag der Zeitschrift f{\"u}r Naturforschung}
}

@article{sugiyama1992static,
  title={Static dielectric response of charged bosons},
  author={Sugiyama, G and Bowen, C and Alder, B J},
  journal={Physical Review B},
  volume={46},
  number={20},
  pages={13042},
  year={1992},
  publisher={APS}
}

@article{sun2015strongly,
  title={Strongly constrained and appropriately normed semilocal density functional},
  author={Sun, Jianwei and Ruzsinszky, Adrienn and Perdew, John P},
  journal={Physical Review Letters},
  volume={115},
  number={3},
  pages={036402},
  year={2015},
  publisher={APS}
}

@article{svensson2024modeling,
  title={Modeling of warm dense hydrogen via explicit real-time electron dynamics: {Dynamic} structure factors},
  author={Svensson, Pontus and Aziz, Yusuf and Dornheim, Tobias and Azadi, Sam and Hollebon, Patrick and Skelt, Amy and Vinko, Sam M and Gregori, Gianluca},
  journal={Physical Review E},
  volume={110},
  number={5},
  pages={055205},
  year={2024},
  publisher={APS}
}

@article{svensson2025modeling,
  title={Modeling of warm dense hydrogen via explicit real-time electron dynamics: {Electron} transport properties},
  author={Svensson, Pontus and Hollebon, Patrick and Plummer, Daniel and Vinko, Sam M and Gregori, Gianluca},
  journal={Physical Review E},
  volume={111},
  number={4},
  pages={045208},
  year={2025},
  publisher={APS}
}

@article{svensson2025accelerated,
  title={Accelerated free energy estimation in \textit{ab initio} path integral {Monte} {Carlo} simulations},
  author={Svensson, Pontus and Kalkavouras, Fotios and Hernandez Acosta, Uwe and Moldabekov, Zhandos A and Tolias, Panagiotis and Vorberger, Jan and Dornheim, Tobias},
  journal={The Journal of Physical Chemistry Letters},
  volume={16},
  number={41},
  pages={10639--10646},
  year={2025},
  publisher={ACS Publications}
}

@article{takada2016emergence,
  title={Emergence of an excitonic collective mode in the dilute electron gas},
  author={Takada, Yasutami},
  journal={Physical Review B},
  volume={94},
  number={24},
  pages={245106},
  year={2016},
  publisher={APS}
}

@article{tao2008nonempirical,
  title={Nonempirical density functionals investigated for jellium: {Spin-polarized} surfaces, spherical clusters, and bulk linear response},
  author={Tao, Jianmin and Perdew, John P and Almeida, Luis Miguel and Fiolhais, Carlos and K{\"u}mmel, Stephan},
  journal={Physical Review B},
  volume={77},
  number={24},
  pages={245107},
  year={2008},
  publisher={APS}
}

@article{tolias2023unravelling,
  title={Unravelling the nonlinear ideal density response of many-body systems},
  author={Tolias, Panagiotis and Dornheim, Tobias and Moldabekov, Zhandos A and Vorberger, Jan},
  journal={Europhysics Letters},
  volume={142},
  number={4},
  pages={44001},
  year={2023},
  publisher={IOP Publishing}
}

@article{troyer2005computational,
  title={Computational Complexity and Fundamental Limitations to Fermionic Quantum {Monte} {Carlo} Simulations},
  author={Troyer, Matthias and Wiese, Uwe-Jens},
  journal={Physical Review Letters},
  volume={94},
  number={17},
  pages={170201},
  year={2005},
  publisher={APS}
}

@article{vorberger2025green,
  title={Green’s Function Perspective on the Nonlinear Density Response of Quantum Many-Body Systems: J. Vorberger et al.},
  author={Vorberger, Jan and Dornheim, Tobias and B{\"o}hme, Maximilian P and Moldabekov, Zhandos A and Tolias, Panagiotis},
  journal={Journal of Statistical Physics},
  volume={192},
  number={6},
  pages={75},
  year={2025},
  publisher={Springer}
}

@misc{vorberger2025roadmapwarmdensematter,
      title={Roadmap for warm dense matter physics}, 
      author={Jan Vorberger and Frank Graziani and David Riley and Andrew D. Baczewski and Isabelle Baraffe and Mandy Bethkenhagen and Simon Blouin and Maximilian P. Böhme and Michael Bonitz and Michael Bussmann and Alexis Casner and Witold Cayzac and Peter Celliers and Gilles Chabrier and Nicolas Chamel and Dave Chapman and Mohan Chen and Jean Clérouin and Gilbert Collins and Federica Coppari and Tilo Döppner and Tobias Dornheim and Luke B. Fletcher and Dirk O. Gericke and Siegfried Glenzer and Alexander F. Goncharov and Gianluca Gregori and Sebastien Hamel and Stephanie B. Hansen and Nicholas J. Hartley and Suxing Hu and Omar A. Hurricane and Valentin V. Karasiev and Joshua J. Kas and Brendan Kettle and Thomas Kluge and Marcus D. Knudson and Alina Kononov and Zuzana Konôpkov á and Dominik Kraus and Andrea Kritcher and Sophia Malko and Gérard Massacrier and Burkhard Militzer and Zhandos A. Moldabekov and Michael S. Murillo and Bob Nagler and Nadine Nettelmann and Paul Neumayer and Benjamin K. Ofori-Okai and Ivan I. Oleynik and Martin Preising and Aurora Pribram-Jones and Tlekkabul Ramazanov and Alessandra Ravasio and Ronald Redmer and Baerbel Rethfeld and Alex P. L. Robinson and Gerd Röpke and François Soubiran and Charles E. Starrett and Gerd Steinle-Neumann and Phillip A. Sterne and Shigenori Tanaka and Aidan P. Thompson and Samuel B. Trickey and Tommaso Vinci and Sam M. Vinko and Lei Wang and Alexander J. White and Thomas G. White and Ulf Zastrau and Eva Zurek and Panagiotis Tolias},
      year={2025},
      eprint={2505.02494},
      archivePrefix={arXiv},
      primaryClass={physics.plasm-ph},
      url={https://arxiv.org/abs/2505.02494}, 
      urldate={2025-08-28},
      note={(accessed 2025-08-28)},
      doi={10.48550/arXiv.2505.02494}
}

@article{wang1992kinetic,
  title={Kinetic-energy functional of the electron density},
  author={Wang, Lin-Wang and Teter, Michael P},
  journal={Physical Review B},
  volume={45},
  number={23},
  pages={13196},
  year={1992},
  publisher={APS}
}

@article{weiss2005path,
  title={Path-integral {Monte} {Carlo} simulations for interacting few-electron quantum dots with spin-orbit coupling},
  author={Weiss, Stephan and Egger, R},
  journal={Physical Review B},
  volume={72},
  number={24},
  pages={245301},
  year={2005},
  publisher={APS}
}

@article{xu2019nonlocal,
  title={Nonlocal kinetic energy density functional via line integrals and its application to orbital-free density functional theory},
  author={Xu, Qiang and Wang, Yanchao and Ma, Yanming},
  journal={Physical Review B},
  volume={100},
  number={20},
  pages={205132},
  year={2019},
  publisher={APS}
}

@article{zastrau2021high,
    author = {Zastrau, Ulf and Appel, Karen and Baehtz, Carsten and Baehr, Oliver and Batchelor, Lewis and Bergh{\"{a}}user, Andreas and Banjafar, Mohammadreza and Brambrink, Erik and Cerantola, Valerio and Cowan, Thomas E and Damker, Horst and Dietrich, Steffen and {Di Dio Cafiso}, Samuele and Dreyer, J{\"{o}}rn and Engel, Hans-Olaf and Feldmann, Thomas and Findeisen, Stefan and Foese, Manon and Fulla-Marsa, Daniel and G{\"{o}}de, Sebastian and Hassan, Mohammed and Hauser, Jens and Herrmannsd{\"{o}}rfer, Thomas and H{\"{o}}ppner, Hauke and Kaa, Johannes and Kaever, Peter and Kn{\"{o}}fel, Klaus and Kon{\^{o}}pkov{\'{a}}, Zuzana and {Laso Garc{\'{i}}a}, Alejandro and Liermann, Hanns-Peter and Mainberger, Jona and Makita, Mikako and Martens, Eike-Christian and McBride, Emma E. and M{\"{o}}ller, Dominik and Nakatsutsumi, Motoaki and Pelka, Alexander and Plueckthun, Christian and Prescher, Clemens and Preston, Thomas R and R{\"{o}}per, Michael and Schmidt, Andreas and Seidel, Wolfgang and Schwinkendorf, Jan-Patrick and Schoelmerich, Markus O. and Schramm, Ulrich and Schropp, Andreas and Strohm, Cornelius and Sukharnikov, Konstantin and Talkovski, Peter and Thorpe, Ian and Toncian, Monika and Toncian, Toma and Wollenweber, Lennart and Yamamoto, Shingo and Tschentscher, Thomas},
    doi = {10.1107/s1600577521007335},
    issn = {1600-5775},
    journal = {Journal of Synchrotron Radiation},
    number = {5},
    pages = {1393--1416},
    title = {The High Energy Density Scientific Instrument at the {European XFEL}},
    volume = {28},
    year = {2021}
}

\end{document}